\documentclass[11pt]{article}
\usepackage{amssymb,amsmath}
\usepackage{bm} 
\usepackage{booktabs} 
\usepackage{array}
\usepackage{latexsym}
\usepackage{graphicx}
\usepackage{color}
\usepackage{datetime}
\usepackage[nosort]{cite}
\usepackage{verbatim}
\usepackage{enumerate}
\usepackage{chngpage} 

\usepackage{psfrag}

\usepackage{mciteplus}

\usepackage[colorlinks=true,      linkcolor=blue,      urlcolor=blue,      
            filecolor=blue,      citecolor=blue,       pdfstartview=FitH,     
						pdfpagemode=UseNone,      bookmarksopen=true]{hyperref}  
\usepackage[all]{hypcap}     

\def\eq#1{(\ref{#1})}

\newcommand{\comm}[1]{} 

\newcommand{\bra}[1]{\langle#1|}
\newcommand{\ket}[1]{|#1 \rangle}

\def\({\left(}
\def\){\right)}
\def\[{\left[}
\def\]{\right]}

\def\lstr{\ell_{\rm str}}
\def\gstr{g_{\rm s}}

\def\sst{\scriptscriptstyle}

\def\coeff#1#2{{\textstyle \frac{#1}{#2}}}

\def\One{{\hbox{ 1\kern-.8mm l}}}

\def\barray{\begin{array}}
\def\earray{\end{array}}
\def\be{\begin{equation}}
\def\ee{\end{equation}}
\def\bea{\begin{eqnarray}}
\def\eea{\end{eqnarray}}
\def\bal{\begin{align}}
\def\eal{\end{align}}
\def\nn{\nonumber}

\def\alphabar{\dot\alpha}

\def\nt{\tilde n}

\numberwithin{equation}{section} 

\makeatletter
\g@addto@macro\bfseries{\boldmath}
\makeatother

\definecolor{cardinal}{rgb}{0.6,0,0}
\definecolor{darkgreen}{rgb}{0,0.4,0}
\definecolor{golden}{rgb}{0.92, 0.7, 0}
\definecolor{midnight}{rgb}{0, 0, 0.5}
\definecolor{darkblue}{rgb}{0, 0, 0.7}

\def\Neql#1{{\cal N}\!=\!{#1}}
\def\Nb{\overline{N}}

\def\IS{\mathbb{S}}
\def\IT{\mathbb{T}}
\def\IZ{\mathbb{Z}}

\def\cA{{\cal A}}
\def\cB{{\cal B}}

\def\cF{{\cal F}}
\def\cG{{\cal G}}
\def\cH{{\cal H}}

\def\cL{{\cal L}}
\def\cM{{\cal M}}
\def\cN{{\cal N}}
\def\cP{{\cal P}}

\def\cR{{\cal R}}
\def\cS{{\cal S}}

\def\cX{{\cal X}}

\def\nBPS#1{$\frac{1}{#1}$-BPS}

\def\p{\partial}

\def\vh{\hat{v}}

\def\bbR{\mathbb{R}}
\def\bbS{\mathbb{S}}
\def\bbT{\mathbb{T}}

\def\ket#1{|{#1}\rangle}

\def\sst#1{\scriptscriptstyle{#1}}

\def\btz{\mathrm{\sst{BTZ}}}

\def\lads{\ell_{\mathrm{\sst{AdS}}}}

\newcommand{\adstwo}{AdS$_2$}
\newcommand{\adsthree}{AdS$_3$}

\newcommand{\qp}{\ensuremath{Q_{\mathrm{P}}}}

\begin{document}

\phantom{AAA}
\vspace{-10mm}

\begin{flushright}
IPHT-T17/135\\
QMUL-PH-17-26\\
YITP-17-127\\
\end{flushright}

\vspace{-2mm}

\begin{center}

{\huge {\bf Asymptotically-flat supergravity solutions \vspace{3mm}\\
 deep inside the black-hole regime}}

\vspace{9mm}

{\large
\textsc{Iosif Bena$^1$,~ Stefano Giusto$^{2,3}$,~ Emil J. Martinec$^{4}$,~ Rodolfo Russo$^{5}$, \vspace{2mm}\\ Masaki Shigemori$^{5,6}$,  David Turton$^{7}$,~ Nicholas P.~Warner$^{8}$}}

\vspace{8mm}

$^1$Institut de Physique Th\'eorique,
Universit\'e Paris Saclay,\\
CEA, CNRS, F-91191 Gif sur Yvette, France \\
\medskip
$^2$Universit\`a di Padova, Via Marzolo 8, 35131 Padova, Italy\\
\medskip
$^3$I.N.F.N. Sezione di Padova, Via Marzolo 8, 35131 Padova, Italy\\
\medskip
$^4$Enrico Fermi Inst.\ and Dept.\ of Physics, \\
University of Chicago,  5640 S. Ellis Ave.,
Chicago, IL 60637-1433, USA\\
\medskip
$^5$Centre for Research in String Theory, School of Physics and Astronomy,\\
Queen Mary University of London, Mile End Road, London, E1 4NS, United Kingdom\\
\medskip
$^6$Center for Gravitational Physics,
Yukawa Institute for Theoretical Physics, \\ Kyoto University,
Kitashirakawa-Oiwakecho, Sakyo-ku, Kyoto 606-8502 Japan\\
\medskip
$^7$Mathematical Sciences and STAG Research Centre, University of Southampton,\\ Highfield,
Southampton SO17 1BJ, United Kingdom\\
\medskip
$^8$Department of Physics and Astronomy
and Department of Mathematics,\\
University of Southern California,
Los Angeles, CA 90089, USA

\vspace{4mm} 
{\footnotesize\upshape\ttfamily iosif.bena @ ipht.fr, stefano.giusto @ unipd.it, ejmartin @ uchicago.edu, r.russo @ qmul.ac.uk,} \\ 
{\footnotesize\upshape\ttfamily  m.shigemori @ qmul.ac.uk, d.j.turton @ soton.ac.uk, warner @ usc.edu} \\

\vspace{11mm}
 
\textsc{Abstract}

\end{center}

\begin{adjustwidth}{12mm}{12mm} 
 
\vspace{-2mm}
\noindent
We construct an infinite family of smooth asymptotically-flat supergravity solutions that have the same charges and angular momenta as general supersymmetric D1-D5-P black holes, but have no horizon. These solutions resemble the corresponding black hole to arbitrary accuracy outside of the horizon: they have asymptotically flat regions, AdS$_3 \times \mathbb{S}^3 $ throats and very-near-horizon AdS$_2$ throats, which however end in a smooth cap rather than an event horizon. The angular momenta of the solutions are general, and in particular can take arbitrarily small values. Upon taking the AdS$_3 \times \mathbb{S}^3$ decoupling limit, we identify the holographically-dual CFT states.

\end{adjustwidth}

\thispagestyle{empty}
\newpage


\baselineskip=14.5pt
\parskip=3pt

\setcounter{tocdepth}{2}
\tableofcontents

\baselineskip=15pt
\parskip=3pt


\section{Introduction and Discussion}
\label{Sect:introduction}

\subsection{An overview of black-hole microstates}

The realization that black holes are thermodynamic black bodies has reshaped our fundamental concept of space and time by introducing profound connections between gravity, quantum mechanics, statistical mechanics and quantum information theory.  The need for a dramatic reformulation of our understanding of horizon-scale physics follows from the fundamental conflict between the locality, causality, and unitarity properties of quantum field theory in the context of black-body (Hawking) radiation emitted by a black hole as described in General Relativity.   Over the years, there has been much debate as to which fundamental physical principles need to be relaxed in order to formulate a consistent theory of quantum gravity.   Investigation of the entanglement structure of Hawking quanta \cite{Mathur:2009hf,Almheiri:2012rt} has sharpened these issues substantially, showing that one cannot simply use effective quantum field theory in the vicinity of a black hole event horizon.  

Gauge/gravity duality~\cite{Maldacena:1997re} strongly suggests that unitarity must survive as a core principle, at least for the class of examples encompassed by this duality.  This is because the space-times on the gravity side of the duality have a time-like boundary and the dual field theories have a standard unitary quantum-mechanical evolution governed by the Hamiltonian conjugate to the preferred global time coordinate on the boundary.

The entire framework of statistical mechanics suggests that the thermodynamic entropy of black holes must be reflected in the statistics of microstate structure.  For theories with a gauge/gravity dual, the underlying density of states is that of the quantum Hilbert space.  The question then arises as to where and how these microstates are encoded in a black hole.   What is the new space-time structure that must emerge at the horizon scale in order to describe a typical black hole microstate?  There are many proposals, ranging from fuzzballs  \cite{Mathur:2005zp,Mathur:2009hf}, firewalls  \cite{Almheiri:2012rt}, Bose-Einstein condensates of gravitons \cite{Dvali:2012wq}, webs of wormholes \cite{Maldacena:2013xja}  or that the information could be encoded in soft photons around the horizon \cite{Hawking:2016msc}.   The problem is that, with the exception of the fuzzball proposal, none of these proposals has a mechanism that is capable of supporting horizon-scale structure against its rapid and inevitable collapse into the black hole. 

The fuzzball proposal, and its developments in the microstate geometry programme, replace the horizons of black holes by higher-dimensional, horizonless  structures that emerge naturally within string theory. The insistence on horizonless structures comes from requiring that quantum unitarity be preserved \cite{Mathur:2009hf}. In terms of the detailed physics, the fuzzball paradigm is that some new phase of matter must emerge at the horizon scale and prevent the formation of the horizon in the first place.  The microstate structure that underpins the black-hole entropy must then remain accessible to outside observers.

The fuzzball programme contains a broad range of distinct enterprises and so two of the authors of this paper proposed the following nomenclature \cite{Bena:2013dka} to refine the relevant ideas:
\begin{enumerate}
\item[1.] A \emph{Microstate Geometry} is a smooth, horizonless solution
	  of supergravity that is valid within the supergravity
	  approximation to string theory and that has the same mass,
	  charge and angular momentum as a given black hole.

\item[2.] A \emph{Microstate Solution} is a formal solution of
	  supergravity equations of motion that is horizonless and that
	  has the same mass, charge and angular momentum as a given
	  black hole.  Microstate solutions are allowed to have
	  Planck/string-scale curvatures corresponding to physical brane sources; non-geometric solutions that can be
	  patch-wise dualized into a smooth solution are also included.
\item[3.] A \emph{Fuzzball} is the most generic horizonless
	  configuration in string theory that has the same mass, charge
	  and angular momentum as a given black hole.  It can involve
	  arbitrary excitations of non-supergravity fields corresponding
	  to massive stringy modes and can be arbitrarily quantum.
\end{enumerate}

Microstate geometries, the first category of microstates above, have
been shown to embody the only semi-classical gravitational
mechanism known thus far that can support horizon-scale microstructure
\cite{Gibbons:2013tqa}. From the perspective of holographic field theory, microstate geometries are intended to capture the infra-red physics of the new
phases of matter that emerge at the horizon scale. Thus, one can argue more generally that the effectiveness of microstate geometries is closely linked to the effectiveness and accuracy of  semi-classical descriptions of
holographic field theory.

Building on the work of~\cite{Mathur:2005zp,Bena:2007kg}, a growing
variety of examples of such geometries have been constructed.  These come in two main classes: ``bubbled geometries'' where all the charges are sourced by Chern-Simons interactions of fluxes threading topology~\cite{Giusto:2004id,Bena:2005va,Berglund:2005vb,Bena:2006kb,Bena:2007qc} (see more recently~\cite{Heidmann:2017cxt, Bena:2017fvm, Avila:2017pwi}); and those in which one of the charges arises from a momentum wave on a bubbled geometry 
\cite{Lunin:2012gp,Giusto:2013bda,Bena:2015bea,Bena:2016agb,Bena:2017geu}.  This
work culminated in some recent key examples outlined in
\cite{Bena:2016ypk}, the details of which we present, and then
generalize, in this paper.

The examples of microstate geometries constructed to date are still
rather limited, and it is not clear whether the most general
configurations are sufficiently generic to represent typical microstates
of a black hole. 
They correspond to macroscopic, coherent excitations of a particular set of modes in the supergravity approximation.
Furthermore, even if there are a macroscopic number of geometric microstates at extremality, it is not clear whether this property will persist far from extremality, although progress in this direction has recently been made~\cite{Bossard:2014ola,Bena:2015drs,Bena:2016dbw,Bossard:2017vii}.

The transition from microstate geometries to the second category~-- microstate solutions~-- 
is expected to encompass more generic horizon-scale microstructure.  For instance,
in the two-charge system in the D1-D5 duality frame, the
microstate geometries involve smooth Kaluza-Klein monopole structures, but
the curvature of the typical configuration lies 
at the scale where the supergravity approximation breaks down
\cite{Martinec:1999sa,Lunin:2002iz} (see also
\cite{Lunin:2002qf,Chen:2014loa}).  
In certain situations adding a third charge has been shown to lower
the curvature and smoothen singular two-charge configurations
\cite{Bena:2016agb}.  Thus it is possible that some portion of the
microstate solutions, once fully backreacted, are actually realized as
microstate geometries.

Microstate solutions also include configurations that are only
patch-wise geometric.  See for example
\cite{Park:2015gka,Fernandez-Melgarejo:2017dme} for attempts to
explicitly construct such microstates, in which different patches of
spacetime are glued together by U-dualities 
\cite{Hull:2004in,Hellerman:2002ax}
and which
might be related to the backreaction of condensates
of stretched branes. \\

Finally, the third, ``Fuzzball'', category is intended to cover the most
general situation that can occur in string theory.
Examples include condensates of stretched branes~\cite{Martinec:2015pfa} that capture a finite fraction of black hole entropy in bubbling microstate geometries~\cite{Bena:2012hf}, and black NS5-branes, whose entropy
can be attributed to the Hagedorn phase of ``little
strings''~\cite{Maldacena:1996ya}. 
However, the proper way to describe
the backreaction of condensates of stretched branes is not yet known.
String theory contains not only massless supergravity fields but also an
infinite tower of massive non-supergravity fields, and it is possible
that they are activated in the most general microstates.  In particular, 
massive stringy modes can be excited very near the horizon~\cite{Giveon:2015cma,Giveon:2016dxe,Silverstein:2014yza,Puhm:2016sxj},
and might distinguish black hole microstates in ways that supergravity cannot.
Furthermore, spacetime itself could become highly quantum, so
that classical geometric notions such as locality and causality might cease to
apply.

The divisions between different categories are not hard and sharp.  For
instance, when curvature is of the order of the string scale, there is no clear-cut
distinction between supergravity modes and stringy modes. In~\cite{Martinec:2014gka} it was argued that fractionation effects
could lead to a geometry which is stringy as seen by some objects and
geometrical as seen by others. Furthermore, one of the authors and Mathur have argued that certain infalling probes interacting
with typical fuzzball microstates may for practical purposes experience a smooth
horizon, for a subset of physical processes~\cite{Mathur:2012jk,Mathur:2013gua}.

The important role of microstate geometries in this overall program is that they represent very explicit, computable examples of geometries that are dual to some of the microstates of black holes.  Moreover, microstate  geometries are capable of supporting extensive microstate structure through classical and semi-classical excitations as well as proving  invaluable  for the study of more ``stringy''  microstate excitations, as in \cite{Martinec:2015pfa}.  Hence, microstate geometries are the laboratory {\it par excellence} for probing and testing ideas about black-hole  microstate structure.

\subsection{Developing the new class of black-hole microstate geometries}

One of the problems inherent in the early constructions of microstate geometries was that all known examples carried angular momenta that are large fractions of the maximally allowed value for the corresponding black holes (see for example~\cite{Bena:2006kb,Bena:2008nh}).  This may have led to a misconception that microstate geometries only exist because of a finely-tuned balance between gravity and angular momentum that keeps the constituent branes spread apart. 
The main mechanism that supports microstate geometries is, in fact, the non-trivial interaction of topological magnetic fluxes.  This enables such geometries to remain macroscopic and non-singular for arbitrarily small angular momentum.  

Typical black-hole microstates should also be  very well-approximated by the black-hole solution until very close to the horizon. For microstate geometries of extremal black holes this requires a long, BTZ-like, \adstwo\ throat.  To obtain such a throat, prior work used bubbling solutions with multiple Gibbons--Hawking (GH) centers  \cite{Bena:2005va,Berglund:2005vb};   the moduli space of these solutions  includes ``scaling'' regions \cite{Denef:2002ru,Bena:2006kb,Bena:2007qc} in which the GH centers approach each other arbitrarily closely, whereupon the solution develops the requisite long \adstwo\ throat.  Quantum effects should set an effective upper bound on the depth of such throats~\cite{Bena:2007qc,deBoer:2008zn}, and a corresponding lower bound on the energy gap, which matches the lowest energy excitations of the (typical sector of the) dual CFT\@.

All the previously-known scaling microstate geometries involve at least three GH centers \cite{Bena:2006kb,Bena:2007qc, Heidmann:2017cxt, Bena:2017fvm, Avila:2017pwi}. Unfortunately, the dual CFT descriptions of these geometries are not yet known.  On the other hand, the holographic dictionary between supergravity solutions and CFT states has been constructed for the generic two-charge states~\cite{Skenderis:2006ah} and for particular three-charge two-centered solutions~\cite{Giusto:2004id,Giusto:2012yz}. Therefore, we were motivated to construct new three-charge black-hole microstate solutions by adding momentum excitations to a certain two-charge, two-center seed solution.  We achieved this using ``superstratum'' technology~\cite{Bena:2014qxa,Bena:2015bea,Bena:2016agb}, which allowed us to introduce momentum-carrying deformations, with specific angular dependence, that modify the momentum and the angular momenta of the solution without introducing new singularities in the geometry  \cite{Bena:2016ypk,Bena:2017geu}. The  geometries in \cite{Bena:2015bea,Bena:2016agb,Bena:2016ypk} were constructed as  excitations of the D1-D5 system in the IIB theory.  The holographic duals of the states were identified as particular left-moving momentum and angular-momentum modes in the D1-D5 CFT \cite{Bena:2015bea,Giusto:2015dfa,Bena:2016agb,Bena:2016ypk}.  In \cite{Bena:2017geu} these results were generalized to M-theory and the MSW string.   

The solutions of this paper depend on several parameters. One parameter lowers the angular momenta, while another parameter adds momentum without increasing the angular momenta of the two-charge seed solution. Thus the angular momentum of the solutions can be parametrically small. These deformations therefore allow us to obtain solutions that have arbitrarily small angular momenta and describe microstates of the non-rotating D1-D5-P (Strominger-Vafa) black hole.  The solutions have an \adstwo\ throat, which becomes longer and longer as the angular momenta $j,\tilde{j} \to 0$, thus classically approximating the non-rotating black hole to arbitrary precision.

\subsection{Near-horizon geometry}

In many examples of holography in five and six dimensions, the
decoupling limit of the near-horizon geometry is asymptotically a sphere
($\IS^2$ or $\IS^3$) fibered over \adsthree, and gravity is dual to a
two-dimensional CFT\@.  In this CFT dual, the asymptotic density of
states is governed by the Cardy formula~\cite{Cardy:1986ie}, for
instance for asymptotically-\adsthree$\times\IS^3$ spacetimes,
\be \label{SCFT} S_{\rm CFT} \;=\; 2\pi\left[ \sqrt{{c\over
6}\left(L_0-{c\over 24}\right) - j^2} + \sqrt{{c\over
6}\left(\tilde{L}_0-{c\over 24}\right) - \tilde{j}^2} \,\right] .  \ee
This formula holds not only for large charges\footnote{Large charges
mean those satisfying $L_0-c/24-6j^2/c\gg c/6.$}
\cite{Hartman:2014oaa}; it remains accurate down to the \emph{cosmic
censorship bound} where $S_{\rm CFT}$ vanishes.  At the bound, the naive
black hole becomes singular; below the bound, the geometries can have
explicit brane sources, or remain smooth and supported by fluxes on
topological cycles, or can have a combination of both.

\begin{figure}[h]
\centerline{\includegraphics[width=3.5in]{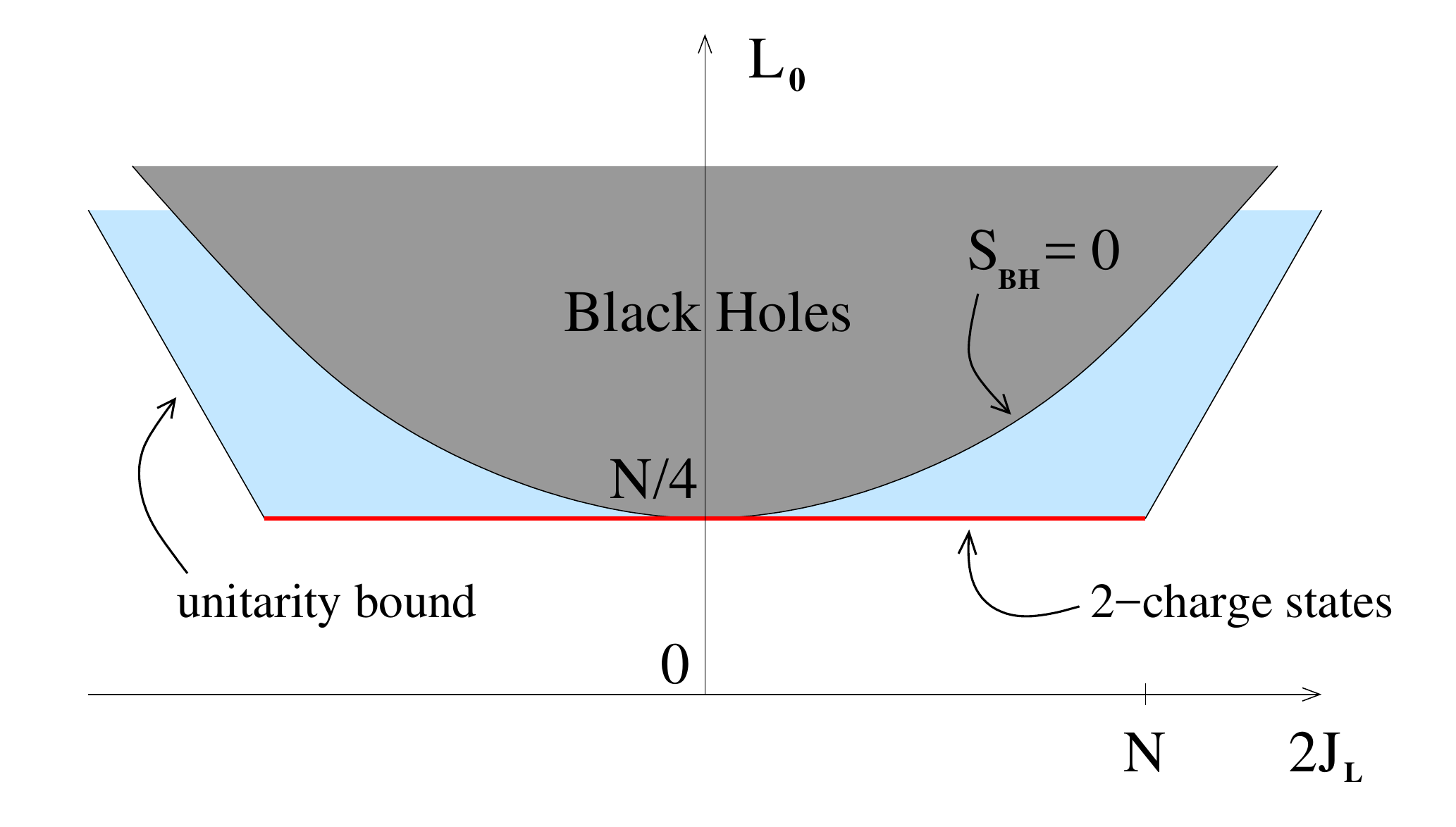}}
\setlength{\unitlength}{0.1\columnwidth} \caption{\it Phase diagram for
the spectrum in the RR sector; note that $c=6N$.  Generic states above the cosmic
censorship bound, i.e.~with $\frac{c}{6} (L_0-\tfrac{c}{24}) \gg  j_L^2$, are microstates with the same
charges as a black hole with rotation on the $\IS^3$; states below
this bound (depicted in blue) are not.  The 1/2-BPS supertube states (on
the red line) all lie at or below the bound.  } \label{fig:Spectrum}
\end{figure}

In this paper we will focus on BPS black holes and so we will only be concerned with the first term in  (\ref{SCFT}).   The na\"{i}ve phase diagram is depicted in Figure~\ref{fig:Spectrum} and the parabola at the boundary of the black hole region is the cosmic censorship bound.
%
The Cardy formula indeed shows that increasing the angular momentum takes away from the free energy available to generate black hole entropy, and takes one closer to a solution with a naked singularity.

The phase diagram of Figure~\ref{fig:Spectrum} is also an
oversimplification.  Near the cosmic censorship bound, there can be a
rich variety of phases involving black holes with other horizon
topologies: for instance one can have black holes localized in both
\adsthree\ and the sphere~\cite{Banks:1998dd,Martinec:1999sa}; black
holes with supertubes around them; three-charge black
rings~\cite{Bena:2004de,Elvang:2004ds,Gauntlett:2004qy}; and
multicenter solutions involving black holes, black rings, and
supertubes~\cite{Bena:2011zw, Crichigno:2016lac}.

To avoid the complications of the phase diagram near extremality with macroscopic angular momentum, and get deep into the black-hole regime, one would  like to be able to dial the angular momentum to small values, while maintaining a large energy above the ground state, so that the corresponding black hole has a macroscopic horizon area.   This was another motivation for constructing the new black-hole microstate solutions outlined in \cite{Bena:2016ypk}.

In six dimensions, the near-horizon geometry of a supersymmetric rotating black string is $\IS^3$ fibered over the extremal BTZ black hole~\cite{Banados:1992wn}, which has the metric\\
\be
\label{dsBTZ}
ds^2_{\btz} =\lads^2\Bigl[\rho^2(-dt^2+dy^2) + \frac{d\rho^2}{\rho^2} + \rho_*^2(dt+dy)^2 \Bigr] \,.
\ee
This metric is locally \adsthree\ and  asymptotes to the standard \adsthree\  form for $\rho \gg \rho_{\ast}$. It can be written as a circle of radius $\rho_{\ast}$ fibered over \adstwo\  in the near-horizon region $\rho \ll \rho_{\ast}$ (see, for example,~\cite{Strominger:1998yg}). Dimensional reduction on this circle yields the \adstwo\ of the near-horizon BMPV solution~\cite{Breckenridge:1996is}. Following the usual abuse of terminology, we will refer to this region as the \adstwo\ throat.

The BTZ parameters and coordinate $\rho$ are related to the supergravity D1, D5, and P charges
$Q_{1}$, $Q_{5}$, $\qp$ and the radial coordinate $r$ (to be used later) as follows. First, we have
 \be
\rho=\frac{r}{\lads^2} ~~,~~~~ \lads^2=\sqrt{Q_1Q_5} ~.  \ee Next, the horizon
radius, $\rho_\ast$, of the extremal BTZ solution \eqref{dsBTZ}
determines the onset of the \adstwo\ throat (and thus the radius of the
fibered $\IS^1$) and is given by \be \rho_\ast^2 = \frac{ \qp}{Q_1Q_5}
~.  \ee This value is determined by a competition between the momentum
charge and the D-brane charges: the former exerts pressure, thereby
expanding the size of the $y$ circle, while the latter exert tension
that tries to shrink the circle.

There has been a growing interest in the physical properties of microstate geometries~\cite{Eperon:2016cdd,Keir:2016azt,Marolf:2016nwu,Eperon:2017bwq} (see also the recent work~\cite{Bena:2017upb,Tyukov:2017uig}). 
In particular, based on a perturbative analysis, it has been argued that supersymmetric microstate geometries are non-linearly unstable when a small amount of energy is added, potentially leading to formation of a black hole~\cite{Eperon:2016cdd}, or an approach to typical microstates~\cite{Marolf:2016nwu}. We note that the asymptotically-flat solutions of this paper break the isometries that were an intrinsic part of the analysis of~\cite{Eperon:2016cdd}, so a more detailed analysis is necessary.
Furthermore, apparently singular behavior can arise when one oversimplifies the system by ignoring degrees of freedom that are necessary for the correct description of the physics.  Therefore the study of these questions requires great care and one must correctly take into account the full phase space of possible configurations explored by the dynamics. The results of this paper advance our understanding of the phase space of microstate geometries. We intend to investigate questions of stability and their physical interpretation in a future work~\cite{instabilities-future}.

\subsection{The structure of this paper}

A brief summary of some of the new microstate geometries that are asymptotic to AdS$_3 \times \IS^3$ appeared in~\cite{Bena:2016ypk}.  In this paper we provide a much more detailed description of their construction, and we generalize these solutions to asymptotically-flat backgrounds.  

We work in type IIB string theory on $\mathbb{R}^{4,1}\times\IS^1\times \cM$, where $\cM$ is $\IT^4$ or $K3$.  The $\IS^1$ is wrapped by $n_1$ D1-branes while $n_5$ D5-branes wrap $\IS^1\times \cM$.  We consider the limit where the volume, $V_4$, of $\cM$ is microscopic and the radius, $R_y$, of the $\IS^1$ (parametrized by the coordinate $y$) is macroscopic, such that the ten-dimensional supergravity brane-charges, $Q_1$ and $Q_5$, are of the same order and macroscopic.  In this limit, the D1-branes and D5-branes provide a heavy background, in which the momentum $P$ along the $y$ direction is a light excitation.  The hierarchy of scales between $R_y$ and $V_4^{1/4}$ means that we can reduce the problem to the low-energy, six-dimensional supergravity theory obtained by reduction on $\cM$.  Following the standard solution-building practice \cite{Bena:2015bea}, we will consider only the supergravity fields that are both independent of the $T^4$, or $K3$, and whose ten-dimensional fields either have no components along $\cM$ or are proportional to the volume form on $\cM$.  The result is six-dimensional $(1,0)$ supergravity coupled to two anti-self-dual tensor multiplets. This system has all the ingredients necessary for the construction of superstrata and has become the workhorse of the microstate geometry programme \cite{Bena:2015bea,Bena:2016ypk, Bena:2017geu}.

Section \ref{Sect:sixD} contains a summary of the six-dimensional supergravity and the equations governing BPS solutions.  These equations can be organized in successive layers. A zeroth layer involves non-linear equations defining the metric of the four-dimensional base space of the solution; in all the solutions in this paper, we will take the same, simple solution for this basic layer. The remaining equations are linear and come in two further layers. The first, which we call Layer~1, is a homogeneous system and, in Section \ref{Sect:first-layer},  we describe how one can find solutions to this system in two-centered geometries using solution-generating techniques. Since this layer of equations is linear, the most general solution can then be obtained from arbitrary superpositions of the simpler solutions obtained by solution-generating methods.  The final layer of BPS equations, which we call Layer~2, is also linear and is sourced by quadratic combinations of the solutions to Layer~1.  In Section \ref{Sect:AdS} we work in a background that is asymptotic to AdS$_3  \times \IS^3$ and solve the final layer of BPS equations when a single mode is excited in Layer 1 of the BPS system.  Such \emph{single-mode superstrata} solutions are structurally much simpler than their multi-mode counterparts~\cite{Bena:2015bea} but illustrate the major points we wish to make here.  In Section \ref{Sect:AsymFlat} we then generalize single-mode superstrata to asymptotically-flat backgrounds. Readers whose interest lies in the new solutions and their properties, rather than in how they are constructed, may wish to skip directly to Sections  \ref{Sect:AdS} and \ref{Sect:AsymFlat}.

Section~\ref{Sect:CFT} contains a review of the structure of the CFT that is dual to string theory on \adsthree$\times\IS^3\times\cM$.  
We identify a particular family of states in the orbifold CFT $\cM^N/S_N$ as the dual to our family of microstate geometries; since the states are BPS, this identification has meaning even though the states being compared lie in completely different loci of the moduli space of the theory.  The Appendices contain some technical details about the supergravity solutions and the normalization of states in the CFT.

The ultimate purpose of this paper is to provide detailed information about the construction of superstrata in both asymptotically-AdS and  asymptotically-flat space-times.  We have provided an extensive introduction so as to set these more technical results in the larger context of the microstate geometry program and we will therefore eschew a conclusions section.  

\section{Supersymmetric D1-D5-P solutions to type IIB supergravity}
\label{Sect:sixD}

As we noted in the previous section, we work in type IIB string theory on $\mathbb{R}^{4,1}\times \bbS^1\times \cM$, where $\cM$ is either $\bbT^4$ or K3. Our solutions are independent of $\cM$, and are described by a six-dimensional $\Neql1$ supergravity coupled to two tensor multiplets. 
The solutions we construct have nontrivial momentum along the common circle wrapped by both the D1 and D5 branes, which is parametrized by $y$ and has radius $R_y$. The first superstrata \cite{Bena:2015bea} were constructed in this theory, which contains all the fields expected from D1-D5-P string emission calculations \cite{Giusto:2011fy}. The system of BPS equations describing all 1/8-BPS D1-D5-P solutions of this theory was derived in  \cite{Giusto:2013rxa}; this is a generalization of the system discussed in \cite{Gutowski:2003rg,Cariglia:2004kk} and simplified in \cite{Bena:2011dd}.

We work with asymptotically null coordinates $u$ and $v$,  related to $y$ and time $t$ via:
\begin{equation}
u ~\equiv~   \frac{1}{\sqrt{2}}\, (t-y) \,,  \qquad v ~\equiv~    \frac{1}{\sqrt{2}}\, (t+y)  \,. \label{uvdefn}
\end{equation}
The BPS solutions have a null isometry along $u$. 

The type IIB ansatz comprises the following ingredients. The six-dimensional metric is a fibration over a four-dimensional base space $\cB$, with metric $ds^2_4$, which may depend on $v$.  
The ansatz includes scalars denoted by $Z_1, Z_2, Z_4, \mathcal{F}$; one-forms on $\cB$ denoted by
$\beta,\omega,a_1, a_2, a_4$;
two-forms on $\cB$ denoted by $\gamma_1,\gamma_2, \delta_2$; and a three-form on
$\cB$ denoted by $x_3$.  All these quantities may depend on $v$ and the coordinates of $\cB$.
These quantities obey BPS equations that we will display momentarily.

We denote the ten-dimensional string-frame metric by $ds^2_{10}$, the six-dimensional Einstein-frame metric by $ds^2_{6}$, the dilaton by $\Phi$, the NS-NS two-form by $B$ and the RR potentials by $C_p$. It is convenient to write $C_6$, the 6-form dual to $C_2$, for the purpose of introducing notation. The full ansatz is~\cite[Appendix E.7]{Giusto:2013rxa}:
\begin{subequations}\label{ansatzSummary}
\allowdisplaybreaks
 \begin{align}
d s^2_{10} &~=~ \sqrt{\alpha} \,ds^2_6 +\sqrt{\frac{Z_1}{Z_2}}\,d \hat{s}^2_{4}\, ,\label{10dmetric}\\
d s^2_{6} &~=~-\frac{2}{\sqrt{\cP}}\,(d v+\beta)\,\Big[d u+\omega + \frac{\mathcal{F}}{2}(d v+\beta)\Big]+\sqrt{\cP}\,d s^2_4\,,\\
e^{2\Phi}&~=~\frac{Z_1^2}{\cP}\, ,\\
B&~=~ -\frac{Z_4}{\cP}\,(d u+\omega) \wedge(d v+\beta)+ a_4 \wedge  (d v+\beta) + \delta_2\,, \label{Bform}\\ 
C_0&~=~\frac{Z_4}{Z_1}\, ,\\
C_2 &~=~ -\frac{Z_2}{\cP}\,(d u+\omega) \wedge(d v+\beta)+ a_1 \wedge  (d v+\beta) + \gamma_2\,,\\ 
C_4 &~=~ \frac{Z_4}{Z_2}\, \widehat{\mathrm{vol}}_{4} - \frac{Z_4}{\cP}\,\gamma_2\wedge (d u+\omega) \wedge(d v+\beta)+x_3\wedge(d v + \beta) \,, \\
C_6 &~=~\widehat{\mathrm{vol}}_{4} \wedge \left[ -\frac{Z_1}{\cP}\,(d u+\omega) \wedge(d v+\beta)+ a_2 \wedge  (d v+\beta) + \gamma_1\right] , 
\end{align}
\end{subequations}
with
\begin{equation}
\alpha \equiv \frac{Z_1 Z_2}{Z_1 Z_2 - Z_4^2}~,~~~~
\cP   \equiv     Z_1 \, Z_2  -  Z_4^2 \,.
\label{Psimp}
\end{equation}
In the above, $d \hat{s}^2_4$ stands for the flat metric on $T^4$, and $\widehat{\mathrm{vol}}_{4}$ denotes the corresponding volume form.

\subsection{The BPS equations}
\label{ss:BPSeqns}

The BPS equations are organized as follows. The four-dimensional metric, $ds^2_4$, and the one-form $\beta$ satisfy non-linear equations; given a solution to this initial set of equations, the remaining ansatz quantities are organized into two layers of linear equations~\cite{Giusto:2013rxa,Bena:2011dd}.

In the current paper we build our solutions within a restricted class of solutions to the non-linear layer of equations, in which the four-dimensional base space is $\mathbb{R}^4$ with $ds_4^2\;\!$ the flat metric, and in which $\beta$ is $v$-independent.
Given this starting point, the BPS equations for $\beta$ simply impose that it has a self-dual field strength,
 \begin{equation}\label{eqbeta}
d \beta = *_4 d\beta\,,
 \end{equation}
where $*_4$ denotes the flat $\mathbb{R}^4$ Hodge dual. 

To write the remaining BPS equations, let us introduce the 2-forms\\
\begin{equation}
\label{Thetadefs}
\Theta_1 ~\equiv~ \mathcal{D} a_1 + \dot{\gamma_2}\,,\quad \Theta_2 ~\equiv~ \mathcal{D} a_2 + \dot{\gamma_1}\,,\quad \Theta_4 ~\equiv~ \mathcal{D} a_4 + \dot{\delta_2}\,.
\end{equation}
Let us denote the exterior differential on the spatial base $\cB$ by $\tilde d$, and introduce 
\begin{equation}
\mathcal{D} \equiv \tilde d - \beta\wedge \frac{\partial}{\partial v}\,.
\end{equation}
The first layer of the BPS equations is then (the dot denotes $\frac{\partial}{\partial v}$):\footnote{The BPS equations \eq{BPSlayer1}, \eq{BPSlayer2} can also be expressed in a covariant form~\cite{Bena:2015bea,Bena:2017geu}.
}
\begin{eqnarray}
 *_4 \mathcal{D} \dot{Z}_1 &=&  \mathcal{D} \Theta_2\,, \qquad \mathcal{D}*_4\mathcal{D}Z_1 ~=~ -\Theta_2\wedge d\beta\,,\qquad \Theta_2~=~*_4 \Theta_2\,,
\cr
 *_4 \mathcal{D} \dot{Z}_2 &=&  \mathcal{D} \Theta_1\,,\qquad \mathcal{D}*_4\mathcal{D}Z_2 ~=~ -\Theta_1\wedge d\beta\,,\qquad \Theta_1~=~*_4 \Theta_1\,,
\label{BPSlayer1} \\
 *_4 \mathcal{D} \dot{Z}_4 &=&  \mathcal{D}  \Theta_4\,,\qquad \mathcal{D}*_4\mathcal{D}Z_4 ~=~ -\Theta_4\wedge d\beta\,,\qquad \Theta_4~=~*_4 \Theta_4\,.
\nonumber
\end{eqnarray}

In (\ref{BPSlayer1}),  the first equation on each line involves four component equations, while the second  equation on each line can be thought of as an integrability condition for the first equation.  The self-duality condition reduces each $ \Theta_I$ to three independent components; including each corresponding equation for $Z_I$ makes four independent functional components, upon which there are four constraints. 

The final set of BPS equations are linear equations for $\omega$ and $\mathcal{F}$, the second of which follows from the $vv$ component of Einstein's equations:
\begin{eqnarray}
\label{BPSlayer2}
\mathcal{D} \omega + *_4  \mathcal{D}\omega +  \mathcal{F}\, d\beta &=&  Z_1 \Theta_1+ Z_2 \Theta_2  -2\,Z_4 \Theta_4 \,, \\
 *_4\mathcal{D} *_4\!\big(\dot{\omega} - \coeff{1}{2}\,\mathcal{D} \mathcal{F} \big) 
 &=& \partial_v^2 (Z_1 Z_2 - {Z}_4^2)  -(\dot{Z}_1\dot{Z}_2  -(\dot{Z}_4)^2 )-\coeff{1}{2} *_4\!\big(\Theta_1\wedge \Theta_2 - \Theta_4 \wedge \Theta_4\big)\,.
\nonumber
\end{eqnarray}
Note that $\mathcal{F}\,d\beta$ appears on the left-hand side of the first equation, so as to separate it from the known sources that arise from the solution to the Layer 1 equations \eq{BPSlayer1}.

\section{First layer of equations: Solution-generating technique}
\label{Sect:first-layer}

In this section we describe the construction of the asymptotically-AdS
solutions, focusing on Layer 1 of the BPS equations. We will discuss
Layer 2 in the next section and the extension of the construction to
asymptotically-flat solutions in Section \ref{Sect:AsymFlat}.

\subsection{The solution-generating technique}
\label{solgen}

Our construction proceeds via the solution-generating technique
developed in \cite{Bena:2015bea}, based on the earlier works
\cite{Mathur:2003hj,Mathur:2011gz,Lunin:2012gp,Giusto:2013bda}. This
technique utilizes the symmetry of the simplest two-charge solution:
after the change of coordinates corresponding to the CFT
spectral flow transformation from the R-R to the NS-NS sector, this solution is
nothing but pure AdS$_3\times \mathbb{S}^3$ and thus has an
$SL(2,\bbR)_L\times SL(2,\bbR)_R\times SU(2)_L\times SU(2)_R$ isometry
group\footnote{This symmetry algebra is enhanced to the full Virasoro
  and current algebras \cite{Brown:1986nw}, but here we do not
  consider them and focus on this ``rigid'' symmetry group.}. One
considers a two-charge solution which is a linear (infinitesimal)
fluctuation around this AdS$_3\times \mathbb{S}^3$ background geometry. We
refer to solutions representing such fluctuations as ``seed
solutions''. If one acts on this linear solution with
$SL(2,\bbR)_L\times SU(2)_L$ generators\footnote{Acting also with the
  right-moving part $SL(2,\bbR)_R\times SU(2)_R$ breaks supersymmetry,
  and is not considered in the present paper. Non-extremal linearized
  solutions where one acts also with $SU(2)_R$ have been recently
  constructed in \cite{Bombini:2017got}. \label{ftnt:no_R_gens}},
then, since the background geometry AdS$_3\times \mathbb{S}^3$ is invariant,
one generates a new linear fluctuation. When written in the original
coordinates describing the R-R states, this fluctuation has a
non-vanishing momentum charge.  The original form of the
solution-generating technique \cite{Mathur:2003hj,Mathur:2011gz,Shigemori:2013lta}
constructed solutions that involve infinitesimal deformations of
AdS$_3\times \mathbb{S}^3$; however we can promote these to solutions
involving finite deformations by using the linear structure of the
BPS equations.

Concretely, we start with a particular two-charge seed solution that has a non-trivial $Z_4$.  The relation of the function $Z_4$
to the profile function defining two-charge solutions is reviewed in
Section \ref{Sect:dual-CFT-states} below; for more details on the
profile function that corresponds to this solution, see
\cite[Eq.~(3.10)]{Bena:2015bea}.

The metric is described in terms of the ansatz quantities described in Section \ref{Sect:sixD} as follows.
The solution has a flat base $\cB=\mathbb{R}^{4}$ which we
write as
 \begin{equation}\label{ds4flat}
 d s^2_4 = (r^2+a^2 \cos^2\theta)\left(\frac{d r^2}{r^2+a^2}+ d\theta^2\right)+(r^2+a^2)\sin^2\theta\,d\phi^2+r^2 \cos^2\theta\,d\psi^2\,.
\end{equation}
Defining 
\begin{equation}
\Sigma \equiv r^2 + a^2 \cos^2\theta\,,
\end{equation}
the expression for the one-form $\beta$ is 
 \begin{equation}\label{eq:beta}
\beta =  \frac{R_y a^2}{\sqrt{2}\,\Sigma}\,(\sin^2\theta\, d\phi - \cos^2\theta\,d\psi) \,.
\end{equation}

We introduce a real parameter $b$; to start with, we consider this to be the amplitude of an infinitesimal fluctuation, and so we allow ourselves to write a complex phase in $Z_4$ for the moment. The functions and forms of the seed solution at linear order in $b$ are as follows
\cite[Eq.~(3.11)]{Bena:2015bea}:
\begin{subequations}
 \label{seed_O(b)}
\begin{align}
 Z_1&= \frac{R_y^2\, a^2}{Q_5 \Sigma},\qquad Z_2 = \frac{Q_5}{\Sigma}\,,
 \qquad \Theta_1=\Theta_2=0,
 \label{seed_O(b)_Z12Th12}
 \\
 Z_4&=R_y\,b\,a^k\,\frac{\sin^k\theta\,e^{-ik\phi}}{(r^2+a^2)^{k/2}\,\Sigma}\,,
 \qquad
 \Theta_4=0,
 \label{seed_O(b)_Z4Th4}
 \\
 \omega &=\frac{R_y\,a^2}{\sqrt{2}\,\Sigma}\,(\sin^2\theta\, d\phi + \cos^2\theta\,d\psi)\equiv \omega_0\,,\qquad
 \mathcal{F}=0\,,
 \label{seed_O(b)_omega_F}
\end{align}
\end{subequations}
where $k$ is a positive integer.  To linear order
in $b$, the relation between the parameters $a,R_y$ and the charges
$Q_1,Q_5$ is
\begin{align}
 a^2 \;=\; {Q_1 Q_5\over R_y^2}\,.
\label{eq:rel_Q1Q5Rya_0}
\end{align}

The ``background geometry'' obtained by setting $b=0$ in the above
solution is global AdS$_3\times \mathbb{S}^3$.  Indeed, in
the new coordinates
\begin{align}
\tilde\phi=\phi -\frac{t}{R_y}
 \,,  \quad 
 \tilde\psi=\psi -\frac{y}{R_y}
 ,
 \label{bulk_spectr_flow}
\end{align}
the six-dimensional metric becomes
\be
ds_6^2 ~=\; \sqrt{Q_1 Q_5}
 \left(- \;\! \frac{r^2+a^2}{a^2 R_y^2}dt^2 +\frac{r^2}{a^2 R_y^2}dy^2 
+\frac{dr^2}{r^2+a^2}
 +d\theta^2+\sin ^2\theta d\tilde\phi^2+\cos ^2\theta d\tilde\psi^2 
 \right),
\label{pure_AdS3xS3}
\ee
which is nothing but global AdS$_3\times \mathbb{S}^3$ with radius $R_{\rm
AdS_3}=R_{\mathbb{S}^3}=\sqrt{a R_y}=(Q_1Q_5)^{1/4}$.  In the dual CFT language, the
coordinate transformation \eqref{bulk_spectr_flow} corresponds to the
spectral flow transformation from the R-R to the NS-NS sector.  We
will refer to the coordinate systems $(t,y,r,\theta,\phi,\psi)$ and
$(t,y,r,\theta,\tilde\phi,\tilde\psi)$ as the R and NS coordinate
systems, respectively.

The generators of the $SL(2,\bbR)_L\times SU(2)_L$ symmetry of AdS$_3\times \mathbb{S}^3$ are
\begin{gather}
\begin{split}
 L_0&={i R_y \over 2}(\partial_t+\partial_y),\\
 L_{\pm 1}
 &=ie^{\pm {i\over R_y}(t+y)}
 \biggl[
 -{R_y\over 2}\biggl({r\over \sqrt{r^2+a^2}}\partial_t+{\sqrt{r^2+a^2}\over r}\partial_y\biggr)
 \pm {i\over 2}\sqrt{r^2+a^2}\,\partial_r
 \biggr],
\end{split} 
\label{sl(2,R)_gen_NS}
\\
 J_0^3=-{i\over 2}(\partial_{\tilde\phi}+\partial_{\tilde\psi}),\quad
  J_0^\pm ={i\over 2}e^{\pm i(\tilde\phi+\tilde\psi)}
 (\mp i\partial_\theta+\cot\theta\, \partial_{\tilde\phi}-\tan\theta\, \partial_{\tilde\psi}).
\label{su(2)_gen_NS}
\end{gather}
These satisfy the standard algebra relations
\begin{align}
 [L_0,L_{\pm 1}]&=\mp L_{\pm 1},\qquad
 [L_1,L_{-1}]=2L_0,\\
 [J^3_0,J^\pm_0]&=\pm J^{\pm}_0,\qquad
 [J^+_0,J^-_0]=2J^3_0.
\end{align}

The solution-generating technique of \cite{Mathur:2003hj}
adapted to our formulation proceeds as follows: (i) extract the six-dimensional or ten-dimensional 
fields from the ansatz quantities of the seed solution; (ii) rewrite the fields in the NS coordinate system using
\eqref{bulk_spectr_flow}; (iii) act on the fields with the NS generators
\eqref{su(2)_gen_NS} to produce a new linear solution; (iv) use
\eqref{bulk_spectr_flow} again to bring the solution back in the R
coordinate system; and finally (v) recast the six-dimensional or ten-dimensional fields into the form of the ansatz,
 and read off the ansatz quantities.

It is cumbersome but straightforward to carry out this procedure
starting with our seed solution \eqref{seed_O(b)}.  This two-charge solution represents a RR ground state,
which can be mapped by spectral flow to an anti-chiral primary state in the NS sector.  An
anti-chiral primary is annihilated by $J^-_0$ and $L_1$ but generates
new (super)descendant states when acted on by $J^+_0$ and $L_{-1}$.  So, in
step (iii) of the above procedure, we act on the seed solution
\be
  m ~~\text{times with}~~ J^+_0  \qquad \text{and} \qquad 
  n ~~\text{times with}~~ L_{-1}\,,
\ee
where $m \le k$, since the action of $(J^+_0)^k$ produces the chiral primary state which is annihilated by any further action of $J^+_0$.

This procedure results in the following ansatz quantities. First of all,
$d s^2_4$, $\beta$, $Z_{1,2}$, $\Theta_{1,2}$,
$\omega$, and $\cF$, are unchanged at linear order in $b$ from their values
given in \eqref{ds4flat}, \eqref{eq:beta}, \eqref{seed_O(b)_Z12Th12}, \eqref{seed_O(b)_omega_F}. Next,
  $Z_4$ and $\Theta_4$ become:
\begin{subequations}
\label{Z4Th4_solngen_O(b)}
\begin{align}
  Z_4&=b\,R_y\,\frac{\Delta_{k,m,n}}{\Sigma}\, e^{-i\hat{v}_{k,m,n}},
\\
 \Theta_4&=
 -\sqrt{2}\,b\,
 \Delta_{k,m,n}
\biggl[i\left((m+n)\,r\sin\theta +n\left({m\over k}-1\right){\Sigma\over r \sin\theta}  \right)\Omega^{(1)}\notag\\
&\hspace{27ex}
 +m\left({n\over k}+1\right)\Omega^{(2)} +\left({m\over k}-1\right)n\, \Omega^{(3)} \biggr]
 e^{-i\hat{v}_{k,m,n}}\,,
\end{align} 
\end{subequations}
where
\begin{align}
\begin{aligned}
  \Delta_{k,m,n} &\equiv
 \left(\frac{a}{\sqrt{r^2+a^2}}\right)^k
 \left(\frac{r}{\sqrt{r^2+a^2}}\right)^n 
 \cos^{m}\theta \, \sin^{k-m}\theta \,, 
 \\
 \hat{v}_{k,m,n} &\equiv (m+n) \frac{\sqrt{2}\,v}{R_y} + (k-m)\phi - m\psi \,,
\end{aligned} 
\label{Delta_v_kmn_def}
\end{align}
and where $\Omega^{(i)}$ ($i=1,2,3$) are a basis of self-dual 2-forms on
$\mathbb{R}^4$:
\begin{equation}\label{selfdualbasis}
\begin{aligned}
\Omega^{(1)} &\equiv \frac{dr\wedge d\theta}{(r^2+a^2)\cos\theta} + \frac{r\sin\theta}{\Sigma} d\phi\wedge d\psi\,,\\
\Omega^{(2)} &\equiv  \frac{r}{r^2+a^2} dr\wedge d\psi + \tan\theta\, d\theta\wedge d\phi\,,\\
 \Omega^{(3)} &\equiv \frac{dr\wedge d\phi}{r} - \cot\theta\, d\theta\wedge d\psi\,.
\end{aligned}
\end{equation}
One can check that the fields \eqref{Z4Th4_solngen_O(b)} satisfy the
Layer 1 BPS equations \eqref{BPSlayer1}.  The Layer 2 equations \eqref{BPSlayer2} are
trivially satisfied by $\omega = \omega_0$ and $\cF=0$, because the fields $Z_4$, $\Theta_4$ are infinitesimal and hence the source terms on the right hand side of 
\eqref{BPSlayer2} are zero.

Let us make a side remark on the CFT state dual to the above solution, to give the reader some rough intuition. The dual holographic description will be fully fleshed out in Section \ref{Sect:CFT}, where the notation used below will be introduced in full.
In the NS-NS sector, the above solution corresponds to a component of the CFT state of the form
\begin{align}
 (J^+_0)^m (L_{-1})^n \ket{00}_k^{\rm NS}\,,
 \label{L,J_on_sigma_k_NS}
\end{align}
where $\ket{00}_k^{\rm NS}$ represents an anti-chiral primary state
related to $Z_4$.  Spectral-flowed to the RR sector, the above component becomes
\begin{align}
 (J^+_{-1})^m (L_{-1}-J^3_{-1})^n \ket{00}_k^{\rm R}\,.
 \label{L,J_on_sigma_k_R}
\end{align}
In the symmetric orbifold CFT, states generically consist of many strands of different lengths.  The state
$\ket{00}_k^{\rm NS,R}$ corresponds to a single strand of length $k$
and the states \eqref{L,J_on_sigma_k_NS} and \eqref{L,J_on_sigma_k_R}
represent their superdescendants.

\subsection{Solution to the first layer of the BPS equations}
\label{sec:firstlayer}

The linear solutions for fields $(Z_4,\Theta_4)$ with quantum numbers
$(k,m,n)$ in \eqref{Z4Th4_solngen_O(b)}, which were obtained by the
solution-generating technique, satisfy the Layer 1 BPS equations
\eqref{BPSlayer1}.  Because these equations are linear differential
equations, we are free to take an arbitrary linear superposition of the
solution \eqref{Z4Th4_solngen_O(b)}, with different finite coefficients
for different values of $(k,m,n)$.
Therefore, the following represents a very general class of solutions to
the ($Z_4$,$\Theta_4$) first layer of the BPS equations: 
\begin{align}
\label{Z4Th4_solngen}
Z_4 &= \sum\limits_{k,m,n} b_{4}^{k,m,n} z_{k,m,n}\,,
 \qquad
 \Theta_4 = \sum_{k,m,n} b_{4}^{k,m,n} \vartheta_{k,m,n}\,,
\end{align}
where  we have defined the mode functions
\begin{align}
 z_{k,m,n}
 &\equiv R_y {\Delta_{k,m,n}\over \Sigma} \cos\hat{v}_{k,m,n}\,,
 \\
\vartheta_{k,m,n}
 &\equiv -\sqrt{2}\,\Delta_{k,m,n}
 \biggl[\left((m+n)r\sin\theta+n\left({m\over k}-1\right){\Sigma\over r \sin\theta}  \right)\Omega^{(1)}\sin\vh_{k,m,n}\notag\\
 &\hspace{25ex}
 +\left(m\left({n\over k}+1\right)\Omega^{(2)}
 +n\left({m\over k}-1\right)\, \Omega^{(3)} \right)\cos\vh_{k,m,n}\biggr] \,.
\end{align}

In writing \eqref{Z4Th4_solngen}, we have taken the real part of
\eqref{Z4Th4_solngen_O(b)}. The coefficients $b_4^{k,m,n}$ are assumed to
be real. More generally we could include different phases for different
values of $(k,m,n)$, but we do not consider that generalization in this paper.
The zeroth-layer fields, $d s^2_4$ and $\beta$ are 
given by \eqref{ds4flat} and \eqref{eq:beta}.

In the symmetric orbifold CFT, having a linear combination of different modes $(k,m,n)$
corresponds to having multiple strands with different quantum numbers
$(k,m,n)$ at the same time. Schematically, instead of
\eqref{L,J_on_sigma_k_R}, the component of the dual CFT state corresponding to the ($Z_4$, $\Theta_4$) solution \eq{Z4Th4_solngen} is now
\begin{align}
\prod_{k,m,n} \Bigl[(L_{-1}-J^3_{-1})^{n} (J^+_{-1})^{m}
 \ket{00}_{k}^{\rm R}\Bigr]^{N_{k,m,n}},\qquad\quad N_{k,m,n}\propto
 \bigl(b_4^{k,m,n}\bigr)^2\,. 
\label{CFT_state_w_multiple_modes}
\end{align}
The fact that the modes are linear fluctuations around
AdS$_3\times \mathbb{S}^3$ is reflected in the relation $N_{k,m,n}\ll N$, which
means that this is an infinitesimal excitation above the R ground
state.  Although the state \eqref{L,J_on_sigma_k_R} was a
superdescendant of the R ground state $\ket{00}_k^{\rm R}$, the
state \eqref{CFT_state_w_multiple_modes} is generically not a superdescendant of
any R ground state and thus is much more general. We will discuss the
form of the CFT states in more detail when we describe the holographic
interpretation of these solutions in Section \ref{Sect:CFT}.

Since the Layer 1 equations \eqref{BPSlayer1} for $(Z_1,\Theta_2)$ and
$(Z_2,\Theta_1)$ are linear and identical to those for
$(Z_4,\Theta_4)$, we can expand $Z_{1,2}$, $\Theta_{1,2}$ in the same modes.  Therefore, a very general set of the
full Layer 1 fields is given by:
\begin{align}
 \label{eq:ZIThI_gen_infinitesimal}
 \begin{aligned}
 Z_1 &= \frac{Q_1}{\Sigma} + \sum\limits_{k,m,n} b_1^{k,m,n} z_{k,m,n} \,,  \quad &
 Z_2 &= \frac{Q_5}{\Sigma} + \sum\limits_{k,m,n} b_2^{k,m,n} z_{k,m,n}\,,\quad &
 Z_4 & = \sum\limits_{k,m,n} b_4^{k,m,n} z_{k,m,n}\,,
\\
 \Theta_1 &= \sum_{k,m,n} b_{2}^{k,m,n} \vartheta_{k,m,n}\,, &
 \Theta_2 &= \sum_{k,m,n} b_{1}^{k,m,n} \vartheta_{k,m,n}\,, &
 \Theta_4 &=  \sum_{k,m,n} b_{4}^{k,m,n} \vartheta_{k,m,n}\,.
 \end{aligned}
\end{align}
In $Z_1,Z_2$, we have included the zero mode parts ${Q_1\over \Sigma}$,
${Q_5\over \Sigma}$ which correspond to empty AdS$_3\times \mathbb{S}^3$ (note
that $z_{0,0,0}\propto {1\over \Sigma}$). 

Now we re-emphasize the crucial observation made above.  A priori, the fields in 
\eqref{eq:ZIThI_gen_infinitesimal} were obtained assuming that the coefficients
$b_I^{k,m,n}$ are infinitesimal.  However, because the Layer 1 equations
are linear differential equations, even if we make the coefficients
$b_I^{k,m,n}$ \emph{finite}, the fields \eqref{eq:ZIThI_gen_infinitesimal} 
continue to exactly solve the Layer 1 equation when the zeroth-layer
fields, $d s^2_4$ and $\beta$, are assumed to be still given by
\eqref{ds4flat} and \eqref{eq:beta}. So we can promote
$b_I^{k,m,n}$ to be finite parameters and the supergravity configuration 
\eqref{eq:ZIThI_gen_infinitesimal}
represents a \emph{finite} deformation of the empty AdS$_3\times \mathbb{S}^3$
background (as far as Layer 1 is concerned). Of course we can perform
the same generalization on the CFT side and assume that the numbers of
strands, $N_{k,m,n}$ in \eqref{CFT_state_w_multiple_modes}, is of order
$N$. It is then natural to ask whether, for the above finite supergravity deformations, there also exists
a simple relation between the CFT and the supergravity
parameters. This issue can be clarified by means of precision
holography tests on the 3-point correlators, as discussed
in~\cite{Skenderis:2006ah,Kanitscheider:2006zf,Giusto:2015dfa}. In
particular it is straightforward to generalize this holographic
analysis to the new states with $n\not=0$ that are the focus of this
paper. 
More concretely, in Section~\ref{Sect:dual-CFT-states}, we will work out
the holographic dictionary in detail for some concrete examples and show
that the amplitude parameter in $Z_4$ in supergravity, $b_4^{k,m,n}$, is
linearly related to the amplitude parameter in CFT; the explicit relation will be given in \eqref{amplmap}.

Thus linearity, which is a result of supersymmetry, has allowed us to
promote the infinitesimal Layer 1 solution generated in the previous
subsection to a finite Layer 1 solution. Once we make $b_I^{k,m,n}$ finite, the Layer
2 equations \eqref{BPSlayer2} require non-trivial solutions depending quadratically on $b_I^{k,m,n}$.  We must compute the
Layer 2 quantities, $\cF$ and $\omega$, by solving the Layer 2 differential
equations \eqref{BPSlayer2} and by requiring that the resulting spacetime is smooth and
free of closed timelike curves. These conditions provide
constraints on the possible values of the $b^{k,m,n}$. However it can be
quite complicated to make these constraints explicit, since it is
usually not obvious how to eliminate singularities in a supergravity
solution. In addition, the details of this procedure depend on the choice of
the Layer 0 fields.

A straightforward ansatz for the coefficients $b_I^{k,m,n}$ that leads to regular solutions is suggested by the above solution-generating technique, extrapolated to non-linear order \cite{Giusto:2013bda,Bena:2015bea}. A systematic procedure to construct exact smooth solutions where the scalars $Z_I$ have the form \eqref{eq:ZIThI_gen_infinitesimal}, starts from the two-charge seed in \cite[Eq.~(3.11)]{Bena:2015bea}, where one keeps also the terms quadratic in $b$, and acts with a finite $SU(2)_L$ rotation\footnote{Acting with finite $SL(2,\bbR)_L$ transformations generates an infinite number of modes and the resulting solution is less easy to analyze.} by an angle $\chi$. The resulting geometry has a finite number of non-vanishing modes $b_4^{k,m,n}$. All the modes generated by this procedure have $n=0$, $m\le k$, and the $b_4^{k,m,n}$ coefficients are not all independent since they contain only two free parameters $b$ and $\chi$.

One can also observe that this procedure results in $b_2^{k,m,n}=0$ for any $(k,m,n)$, and hence all the $Z_2$ modes are trivial. However, the modes of $Z_1$ are nontrivial, and depend  quadratically on the coefficients $b_4^{k,m,n}$. The relation between the coefficients $b_4^{k,m,n}$ and $b_1^{k,m,n}$ is such that the sources for the second-layer equations \eqref{BPSlayer2} depend only on the difference of the modes $\hat v_{k,m,n}-\hat v_{k',m',n'}$ but not on their sum. The solutions generated in this way are by construction superdescendants of two-charge states and represent only a small subset of the general solutions considered above, where one has modes with arbitrary $k,m,n$ and the coefficients $b_4^{k,m,n}$ are arbitrary. One can however exploit the linearity of the first layer of equations and extrapolate the structure of the coefficients $b_I^{k,m,n}$ found for superdescendants to a generic superposition of modes. This is the ansatz that was taken in \cite{Bena:2015bea} for constructing superstrata with $n=0$, and in the next section we will follow the same approach.

\section{Second Layer of the BPS equations: Asymptotically AdS}
\label{Sect:AdS}

In this section we describe the construction of solutions to  Layer 2 of the BPS equations, focusing on asymptotically-AdS solutions. Asymptotically-flat solutions will be presented in the next section.  However, before we focus on particular asymptotics, we now make some general remarks  outlining some key elements of the structure of the second layer of BPS equations  (\ref{BPSlayer2}) that enable us to break the problem into manageable pieces.  

First, the sources on the right-hand side of (\ref{BPSlayer2}) are quadratic in the $Z$'s and $\Theta$'s, which means that the sources involve the sums and differences of their Fourier mode dependences, $\vh_{k,m,n}$.  Explicitly, there are two types of source: those with phase dependence $\vh_{k+k',m+m',n+n'}$, and those with phase dependence $\vh_{k-k',m-m',n-n'}$ (here we assume that $k-k' \ge 0$ without loss of generality).

As mentioned at the end of the previous section, for superdescendant states one finds no sources with phase $\vh_{k+k',m+m',n+n'}$. Furthermore, based on experience \cite{Bena:2015bea}, when mode dependences $\vh_{k,m,n}$ add together, the corresponding solution to Layer 2  (\ref{BPSlayer2}) is generically singular. 
In this paper we will always arrange that these Layer 2 sources are absent.
Thus our strategy will be to set the $b_2$-modes to be zero, and to tune the $b_1$-coefficients so as to cancel the terms with $\vh_{k+k',m+m',n+n'}$ in the Layer 2 sources. Note that for a pair of modes $(k,m,n)$ and $(k',m',n')$, such a cancellation is not possible if $(k m'-k' m)(k n'-k' n)\not=0$, unless one excites other fields. Thus, if one allows generic modes to interact, the construction of regular solutions could prove rather more challenging.

By adjusting the Fourier coefficients in $(Z_1,\Theta_2)$ in terms of those in $(Z_4,\Theta_4)$ in this way, one can construct fully smooth microstate geometries. This tuning of Fourier coefficients to create a smooth outcome is known as ``coiffuring''  \cite{Mathur:2013nja,Bena:2013ora,Bena:2014rea}. Since the sources of Layer 2 are quadratic in $Z_I$ and $\Theta_I$, the $b_1$ coefficients depend quadratically on the $b_4$ coefficients. 

We emphasise that the Fourier coefficients $b_4^{k,m,n}$ of $Z_4$ are allowed to remain arbitrary, in agreement with the results of IIB string scattering amplitudes \cite{Giusto:2011fy,Giusto:2009qq,Black:2010uq,Bianchi:2016bgx}. We will see that this choice makes the source terms in the Layer 2 equations particularly simple, and leads to smooth solutions. Because of the obstruction mentioned above, this approach is not directly applicable to the most general superposition of modes, depending on both $m$ and $n$. However, interactions between multiple Fourier modes were considered in \cite{Bena:2015bea} for $n=0$ and that approach should work whenever each pair of modes satisfies $(k m'-k' m)(k n'-k' n)=0$. In particular, we expect that the construction of multi-mode solutions with $m=m'=0$ will be possible using methods very similar to those employed in \cite{Bena:2015bea}.

To keep things simple, in this paper we will only construct solutions with a single mode, for which this issue does not arise. 
For a single Fourier mode, there will be terms with phase dependence $\vh_{2k,2m,2n}$, and there will be ``RMS'' modes, proportional to the square of the Fourier coefficient $(b_{4}^{k,m,n})^2$ but independent of $(v,  \phi,\psi)$.  We will deal with each separately.

The non-oscillating RMS terms depend only upon $(r, \theta)$ and the contributions to $\omega$ and $\cF$ from these terms simplify to:
\begin{equation}
\omega^{\rm RMS} ~=~ \omega_1(r,\theta)  \, d\phi ~+~ \omega_2(r,\theta)  \, d\psi\,, \qquad  \cF^{\rm RMS}  =   \cF(r,\theta)   \,.
\label{RMSform}
\end{equation} 
As we will see, these equations can be solved completely, albeit in a form involving sums of multinomial coefficients.  Physically, these RMS parts of the solution contain the longer-distance effects of the oscillations, encoding all the resulting changes (with respect to the seed solution) in the asymptotic momentum charge and angular momenta. 

To solve the equations for  oscillating sources one can use a gauge invariance of (\ref{BPSlayer2}) to set\footnote{Note that this choice is only possible for modes that have a non-trivial $v$-dependence, of the type we will consider in this paper.} 
\begin{equation}
 \cF^{\rm osc} ~=~  0\,.
\label{gaugechoice1}
\end{equation}
Having made this gauge choice, one can write (\ref{BPSlayer2}) in terms of differential operators acting on each component of $\omega$.  From experience  \cite{Bena:2015bea}, one typically finds that  this system can first be reduced to a Laplacian on the sum of components $(\omega_\psi+\omega_\phi)$, and once this equation is solved, with a little guesswork one can leverage this to find the complete solution for all the components of $\omega$.  We will describe this procedure in more detail in Section \ref{Solving-layer2}.

With only a single mode, and for asymptotically-AdS solutions, the coiffuring results in a complete cancellation of the mode dependence in the metric. Hence the metric is completely independent of $(v, \psi,\phi)$.  In these solutions, the tensor fields still oscillate as functions of $(v, \psi,\phi)$, but the coiffuring cancels these oscillations in the energy-momentum tensor and so the gravitational field does not oscillate.  The gravitational field does respond to the fluctuations, but only through their RMS effects.  Thus, the single-mode asymptotically-AdS superstrata which we construct in this section are the simplest of their kind, and their second-layer equations (\ref{BPSlayer2}) have only non-oscillating, RMS sources.  

To obtain asymptotically-flat superstrata, one must ``add $1$'s'' to $Z_1$ and $Z_2$, and this creates new source terms that depend explicitly upon the oscillations in $(v, \psi,\phi)$.  This requires us to find new families of solutions to  (\ref{BPSlayer2}). These solutions will be constructed in Section \ref{Sect:AsymFlat} and, as we will see, their metric will depend non-trivially upon $(v, \psi,\phi)$ even after coiffuring.

\subsection{Solution to the second layer of the BPS equations}
\label{RMSsols}

Following \cite{Bena:2015bea} we set the oscillations in $(Z_2,  \Theta_1)$ to zero, since this choice emerges naturally from the non-linear solution-generating method described at the end of Section \ref{sec:firstlayer}. We also specialize to a single-mode superstratum, which means reducing to single Fourier modes in (\ref{eq:ZIThI_gen_infinitesimal}).  The structure of the quadratic sources in Layer 2 means that it is  natural for the modes of $(Z_1,  \Theta_2)$ to have twice the mode numbers of $(Z_4,  \Theta_4)$.  
Since we now specialize to a single mode, we suppress the ($k,m,n$) indices on $b_I^{k,m,n}$.
Thus  we take the full Layer 1 fields to have the form:
\begin{equation}
\begin{aligned}
Z_1 ~=~ & \frac{Q_1}{\Sigma} + \frac{b_1\, R_y^2 }{2 Q_5} \,
\frac{\Delta_{2k,2m,2n}}{\Sigma} \cos \vh_{2k,2m,2n} \,,   \qquad Z_2 ~=~ \frac{Q_5}{\Sigma} \,,  \\
Z_4~=~   & R_y  \, b_4\, \frac{ \Delta_{k,m,n} }{\Sigma} \cos \vh_{k,m,n}  \,,
\end{aligned}
 \label{eq:ZIAdSsinglemode} 
\end{equation}
with
\begin{equation}
\label{eq:ThetaIAdSsinglemode}
 \Theta_1 =0\,,\qquad
 \Theta_2 = \frac{b_1 R_y}{2 Q_5}\, \vartheta_{2k,2m,2n}\,,\qquad
 \Theta_4 = b_4\, \vartheta_{k,m,n}\,.
\end{equation}

With these choices, the sources of the Layer 2 BPS equations have  an oscillating part that depends only upon $\vh_{2k,2m,2n}$ as  well as an RMS part.   As in \cite{Bena:2015bea}, we find that such oscillating sources generically lead to singular angular momentum vectors, $\omega$. However, the Fourier coefficient of the oscillating source is proportional to $b_1 - b_4^2$ and so we take:
\begin{equation} 
b_1 ~=~b_4^2\,.
\label{AdScoiff}
\end{equation} 
This coiffuring of the modes removes the singular oscillating  parts and leaves us with only the RMS sources.  As we will see, this leads to a smooth solution.

The solution for $\omega$ and $\cF$ is now given by the sums of the original supertube solutions and the solution for the RMS pieces, as in  (\ref{seed_O(b)_omega_F}) and (\ref{RMSform}):
\begin{equation} 
\omega^{\rm AdS} ~=~ \omega_0  ~+~ \omega^{\rm RMS}\,,  \qquad  \cF ~=~ \cF^{\rm RMS}\,.
\label{AdSomega}
\end{equation} 
The equations (\ref{BPSlayer2}) for $\omega^{\rm RMS}$ now reduce to: 
\bea
d \omega^{\rm RMS} + *_4 d \omega^{\rm RMS} + \mathcal{F} \,d\beta
&=& 
\sqrt{2}\,R_y\,b_{4}^2 \frac{\Delta_{2k,2m,2n}}{\Sigma} \left(
\frac{m(k+n)}{k} \,\Omega^{(2)} - \frac{n(k-m)}{k} \,\Omega^{(3)} \right) , 
\cr \label{eq:omegakmn}&&
\eea
\bea \label{eq:Fkmn}
\widehat{\mathcal{L}} \,  \mathcal{F}
&=&
\frac{4 \;\! b_{4}^2}{r^2+a^2}\frac{1}{\cos^2\theta \, \Sigma} 
\left[ \left( \frac{m(k+n)}{k} \right)^{2} \Delta_{2k,2m,2n} 
+ \left( \frac{n(k-m)}{k} \right)^{2} \Delta_{2k,2m+2,2n-2} \right] , \cr &&
\eea
where $\widehat{\mathcal{L}}$ is the scalar Laplacian on the base space $\cB$:
\begin{equation}
\widehat{\mathcal{L}} F \equiv \frac{1}{r\Sigma}\, \partial_r \big( r (r^2 + a^2) \, \partial_r F  \big)  +    \frac{1}{\Sigma\sin \theta \cos \theta}\partial_\theta \big( \sin \theta \cos \theta\, \partial_\theta F  \big)\,.
\end{equation}

Since the right-hand side of \eq{eq:omegakmn} has no component in the $\Omega^{(1)}$ direction, we can set the components $\omega_r = \omega_{\theta} = 0$. We write
\begin{equation}
\omega^{\rm RMS}  \;\equiv\;  \mu_{k,m,n} (d\psi+d\phi) + \zeta_{k,m,n}(d\psi-d\phi)\,.
\label{eq:omkmnparts1}
\end{equation}
Inspired by the results of \cite{Niehoff:2013kia,Bena:2015bea}, we define
\begin{equation}
\hat \mu_{k,m,n} \equiv \mu_{k,m,n} +\frac{R_y}{4\sqrt{2}}\frac{r^2+a^2\sin^2\theta}{\Sigma}\mathcal{F}_{k,m,n}+\frac{R_y\,b_4^2}{4\sqrt{2}} \,\frac{\Delta_{2k,2m,2n}}{\Sigma}\,,
\end{equation}
where $\mathcal{F}_{k,m,n}\equiv \cF$ is the solution of \eqref{eq:Fkmn}.
Then $\hat \mu_{k,m,n}$ satisfies 
\begin{equation}
\widehat{\mathcal{L}}\, \hat \mu_{k,m,n} = 
 \frac{R_y\,b_4^2}{\sqrt{2}}\frac{1}{(r^2+a^2)} \frac{1}{\cos^2\theta \, \Sigma} 
\left(\frac{(k-m)^2(k+n)^2}{k^2} \Delta_{2k,2m+2,2n}
+\frac{(nm)^2}{k^2} \Delta_{2k,2m,2n-2}\right) .
\end{equation} 
Once $\mu_{k,m,n}$ has been computed, $\zeta_{k,m,n}$ is determined by substituting \eq{eq:omkmnparts1} into \eq{eq:omegakmn}, which gives ($s_{\theta}=\sin\theta$, $c_{\theta}=\cos\theta$)
\begin{equation}
\begin{aligned}
\partial_r \zeta_{k,m,n} &= 
\frac{r^2 \cos2\theta-a^2 s_{\theta}^2}{r^2+a^2 s_{\theta}^2} \partial_r\mu_{k,m,n}
 -\frac{r \sin2\theta}{r^2+a^2 s_{\theta}^2}\partial_\theta \mu_{k,m,n}\\
&\quad+\frac{\sqrt{2}R_y\,r}{\Sigma(r^2+a^2 s^2_{\theta})}
\left[ b_4^2 \Bigl( m s^2_{\theta} + n c^2_{\theta} -\frac{mn}{k} \cos 2 \theta \Bigr) \Delta_{2k,2m,2n}-\frac{a^2 (2r^2+a^2)s^2_{\theta}c^2_{\theta}}{\Sigma}\mathcal{F}_{k,m,n}\right] \,,\\
\partial_\theta\zeta_{k,m,n}&=\frac{r(r^2+a^2) \sin2\theta}{r^2+a^2 s_{\theta}^2}\partial_r \mu_{k,m,n}+ \frac{r^2 \cos2\theta-a^2 s_{\theta}^2}{r^2+a^2 s_{\theta}^2} \partial_\theta\mu_{k,m,n}\\
&\quad+\frac{R_y\,\sin 2 \theta}{\sqrt{2}\,\Sigma\,(r^2+a^2 s^2_{\theta})}
\left[ b_4^2 \Bigl( - m r^2+ n (r^2+a^2) - \frac{mn}{k} (2r^2+a^2) \Bigr) \Delta_{2k,2m,2n} \phantom{\frac{r^2}{\Sigma}} \right. \\
&\qquad\qquad\qquad\qquad\qquad\qquad{}
 \left. +\frac{a^2 r^2 (r^2+a^2)\cos 2\theta}{\Sigma}\mathcal{F}_{k,m,n}\right] .
\end{aligned} \label{eq:zeta-eqns}
\end{equation} 

To solve the equations for $\cF$ and $\hat\mu_{k,m,n}$, we must
find the function $F_{2k,2m,2n}$ that solves the equation
\begin{align}
 \widehat{\cL}F_{2k,2m,2n}={\Delta_{2k,2m,2n}\over (r^2+a^2)\cos^2\theta\,\, \Sigma} \;.
\end{align}
In Appendix \ref{app:F_kmn}, we find that the solution to this problem is given by
\begin{equation} 
F_{2k,2m,2n}\,=\,-\!\sum^{j_1+j_2+j_3\le k+n-1}_{j_1,j_2,j_3=0}\!\!{j_1+j_2+j_3 \choose j_1,j_2,j_3}\frac{{k+n-j_1-j_2-j_3-1 \choose k-m-j_1,m-j_2-1,n-j_3}^2}{{k+n-1 \choose k-m,m-1,n}^2}\,
\frac{\Delta_{2(k-j_1-j_2-1),2(m-j_2-1),2(n-j_3)}}{4(k+n)^2(r^2+a^2)}\,,
\end{equation} 
where
\begin{equation} 
{j_1+j_2+j_3 \choose j_1,j_2,j_3}\equiv \frac{(j_1+j_2+j_3)!}{j_1! j_2! j_3!}\,.
\end{equation} 
In terms of $F_{2k,2m,2n}$, the form of $\cF\equiv \cF_{k,m,n} $ and $\mu_{k,m,n}$ for general $k,m,n$ is
\begin{align}
\label{cF}
\cF_{k,m,n} &= 4b_4^2\biggl[\frac{m^2 (k+n)^2}{k^2}\,F_{2k,2m,2n}+\frac{n^2 (k-m)^2}{k^2}\,F_{2k,2m+2,2n-2}\biggr],
 \\
\mu_{k,m,n}&= \frac{R_y\,b_4^2}{\sqrt{2}}\,\biggl[ 
\frac{(k-m)^2(k+n)^2}{k^2} F_{2k,2m+2,2n}
+\frac{m^2 n^2}{k^2} F_{2k,2m,2n-2}
\notag\\
&\hspace{20ex}
 -\frac{r^2+a^2\,\sin^2\theta}{4\,\Sigma}\,b_4^{-2}\mathcal{F}_{k,m,n}
-\frac{\Delta_{2k,2m,2n}}{4\,\Sigma}
+\frac{x_{k,m,n}}{4\,\Sigma}
\biggr]\,.
\label{mu}
\end{align} 
In this expression for $\mathcal{F}_{k,m,n}$ and $\mu_{k,m,n}$ it should
be understood that, when the coefficient of an $F$ function is zero, the
term is zero. The term proportional to $x_{k,m,n}$ is a harmonic piece that we can freely add to the solution of the Poisson equation for $\hat \mu_{k,m,n}$. The coefficient $x_{k,m,n}$ will be fixed by regularity in the next subsection.

At this point, the equations \eq{eq:zeta-eqns} for $\zeta_{k,m,n}$ can be solved by quadrature.

Having obtained the asymptotically-AdS solutions, the conserved charges can be computed from supergravity in order to be compared to the dual CFT states.  However, it is simpler to obtain these charges from the asymptotically-flat solutions that will be constructed in the next section (yielding the same values of the charges), so we postpone the analysis until that point.

\subsection{Regularity}
\label{Sect:regularityAdS}

Regularity of the solution requires that the metric and all other ten-dimensional fields never diverge and that the components of the metric and the forms are well-behaved at points where our coordinate system degenerates. There are potential divergences at the supertube location $\Sigma=0$ $(r=0,\theta=\pi/2)$, which must be taken care of; the smoothness analysis follows the pattern familiar from the study of two-charge supertube solutions. The loci where our coordinate system degenerates are: (i) the plane $\theta=0$, where the $\phi$-cycle shrinks, (ii) the plane $\theta=\pi/2$, where the $\psi$-cycle shrinks, (iii) the point $(r=0,\theta=0)$ where the whole angular $\mathbb{S}^3$ shrinks. Functions of $\phi$ and $\psi$ must vanish sufficiently fast on the planes (i) and (ii) to be smooth: more precisely $e^{\pm i \,m \,\phi}$ ($e^{\pm i \,m \,\psi}$) must vanish at least like $\theta^{m}$ ($(\theta-\pi/2)^m$) for $\theta\to 0$ ($\theta\to \pi/2$). Analogous, but more stringent, conditions apply to forms with legs along $\phi$ and/or $\psi$: the general requirement is that components of forms must be regular when expressed in a well-behaved local orthonormal frame. It is easy to verify that our solutions satisfy these requirements: for example the $\theta$-dependence of the factor $\Delta_{k,m,n}$ guarantees that the function $\Delta_{k,m,n} \,\cos \hat v_{k,m,n}$ is well-behaved on the planes (i) and (ii). The analysis of the point $(r=0,\theta=0)$ requires more care and will be discussed in more detail below. One should also verify that the metric has no CTCs: as is usual, a complete proof valid for the general class of solutions would be complicated, however we can show that the metric is well-behaved in the most dangerous regions $(r=0, \theta =0)$ and $(r=0, \theta =\pi/2)$, and there is no reason to expect problems elsewhere. Furthermore, for the explicit example sub-family of $(k,m,n)=(1,0,n)$ that we shall present, we will prove the absence of CTCs.

\subsubsection{Near $(r=0, \theta=0)$}

The point $(r=0, \theta =0)$ represents the origin of polar coordinates on the flat $\mathbb{R}^4$ base; to analyze the behaviour of the solutions around this point it is convenient to switch to ordinary polar coordinates $(\tilde r, \tilde \theta)$ and take the limit $\tilde r\to 0$ with fixed $\tilde \theta$. In this limit one has
\bea\label{eq:tildertheta}
r \approx \tilde r \cos\tilde \theta\,,\quad \sin\theta \approx \frac{\tilde r \sin\tilde \theta}{a}\,.
\eea
Moreover the one-form $\beta$ introduces a mixing between $v$ and $\psi$ (as can be seen from $dv+\beta\approx d v -R_y/\sqrt{2}\, d\psi$), so that it is convenient to work with the coordinate 
\be\label{eq:vtilde}
\tilde v \equiv v - \frac{R_y}{\sqrt{2}}\,\psi
\ee
around this point. Then the combination that appears in the scalars $Z_4$ and $Z_1$,
\bea
\Delta_{k,m,n} \cos \hat v_{k,m,n} \sim {\tilde r}^{n+k-m} \cos^n\tilde \theta\,\sin^{k-m}\tilde \theta \,\cos\left[(m+n)\frac{\sqrt{2}\,\tilde v}{R_y} + n\,\psi +(k-m)\,\phi \right] ,
\eea
satisfies the criterion for regularity around $\tilde r =0$. For one-forms with legs along $\phi$ and $\psi$, further conditions have to be met. In solutions with a single mode, $\omega$ does not depend on $v$, $\phi$ or $\psi$, so a sufficient condition for regularity is that both the $\phi$ and $\psi$ components of $\omega$ vanish for $\tilde r\to 0$. For the component along $d\phi+d\psi$, we can use the general expression given in (\ref{mu}). Requiring that $\mu_{k,m,n}$ vanishes for $(r=0,\theta=0)$ fixes the value of the constant $x_{k,m,n}$ that was left undetermined in \eqref{mu}:
\bea \label{eq:x}
x_{k,m,n}^{-1}= {k \choose m} {k+n-1 \choose n} \,.
\eea 
As we do not have a general closed-form expression for the $d\phi-d\psi$ component of $\omega$, its vanishing has to be checked case by case: for example this condition is satisfied by the example sub-family of $(k,m,n)=(1,0,n)$ that will be given in (\ref{eq:omega10n}).

\subsubsection{Near $(r=0, \theta=\pi/2)$}

When $(r=0, \theta=\pi/2)$, both the scalars $Z_1$, $Z_2$ and $Z_4$ and the one-forms $\beta$ and $\omega$ diverge, and these divergences must cancel for the metric to be smooth. It turns out to be sufficient to require the cancellation of the divergent part in the $(d\phi+d\psi)^2$ component of the metric. The resulting condition is
\be\label{eq:Q1Q5}
\frac{Q_1 Q_5}{R_y^2} = a^2+ \frac{b^2}{2}\,,\quad b^2 =  x_{k,m,n}\,b_4^2\,,
\ee
with $x_{k,m,n}$ given in (\ref{eq:x}). 
This condition can be thought of as determining the non-oscillating part of $Z_1$, which is proportional to $Q_1$.
All other divergences cancel as a consequence of this condition. 
For solutions with only one mode,  the condition (\ref{eq:Q1Q5}) also ensures that the warp factor $Z_1$ is everywhere positive, no matter how large the amplitude of the fluctuations. Indeed the minimal value of $Z_1$ is attained for $\cos\hat v _{2k,2m,2n}=-1$, and then the identity
\begin{equation} \label{combinatoric1} 
\frac{\Delta_{2k,2m,2n}}{x_{k,m,n}} \le  \sum_{k'=1}^\infty\sum_{n'=0}^{\infty}\sum_{m'=0}^{k'}\delta_{k'+n', k+n}\, \frac{\Delta_{2k',2m',2n'}}{x_{k',m',n'}}  ~=~  \frac{a^2 }{(r^2 +a^2)} \le 1\,
\end{equation} 
guarantees that $\,b_4^2 \;\! \- \Delta_{2k,2m,2n} < b^2$ and hence $Z_1>0$ for all our three-charge solutions.

\subsection{Examples}

The class of solutions with $n=0$ was described in detail in \cite{Bena:2015bea} where several examples were discussed. A family of solutions where all three quantum numbers are non-trivial ($k=2$, $m=1$ and $n$ arbitrary) can be found in~\cite{Bena:2017upb}. Here we focus on the class $m=0$, presenting some examples in closed form and a general algorithm that can be used to generate further solutions.

\subsubsection{The $m=0$ class}
\label{Sect:k1m0sol}

For $k=1$, $m=0$ and generic $n$, one finds
\begin{equation}
\cF_{1,0,n} ~=~ - \frac{b_4^2}{a^2} \, \bigg(1 - \frac{r^{2n}}{(r^2+a^2)^n}\bigg)
\end{equation}
and integrating for $\zeta$ yields $\omega_{1,0,n}$:
\begin{equation}\label{eq:omega10n}
\omega_{1,0,n} ~=~   
 \frac{b_4^2 \, R_y}{\sqrt{2} \, \Sigma}\, \bigg(1 - \frac{r^{2n}}{(r^2+a^2)^n}\bigg) \, \sin^2\theta\, d\phi\,. 
\end{equation}
For $k\!=\!2$, $m\!=\!0$, one finds:
\begin{equation}
\begin{aligned}
\cF_{2,0,n} ~=~ & - \frac{b_4^2}{(n+1)^2\, a^4} \bigg[  n a^2 - r^2\bigg( 1 - \frac{r^{2n}}{(r^2 +a^2)^{n}} \bigg) \\
& \qquad~+~ \bigg( \bigg( 1 - \frac{r^{2n}}{(r^2 +a^2)^{n}} \bigg)(2r^2 + (2n+1) a^2)  - 2 n a^2  - \frac{n^2 a^4 \,r^{2n}}{(r^2 +a^2)^{n+1}} \bigg) 
\sin^2 \theta \bigg]  \,, \\ 
\omega_{2,0,n}~=~ & \frac{ R_y}{\sqrt{2}\,\Sigma}\bigg\{ \,\frac{b_4^2}{(n+1)^2}\, \bigg[ (n+1) \bigg(  1 - \frac{r^{2n}}{(r^2 +a^2)^{n}}  -\frac{ n \, a^2 \, r^{2n}}{(r^2 +a^2)^{n+1}} \bigg) \,   \\
& \qquad  \qquad~-~ \bigg( \frac{r^2}{a^2}\,\bigg( 1 - \frac{r^{2n}}{(r^2 +a^2)^{n}} \bigg) ~-~ n\bigg) \, \cos^2 \theta \bigg] \, \bigg\} \, \,  \sin^2 \theta \,d\phi \\
&~-~\frac{ R_y}{\sqrt{2}\,\Sigma}\, \frac{b_4^2}{(n+1)^2}\, \bigg[ \frac{r^2}{a^2}\,\bigg( 1 - \frac{r^{2n}}{(r^2 +a^2)^{n}} \bigg) -  \frac{n \, r^{2n+2}}{(r^2 +a^2)^{n+1}} \bigg]\, \sin^2 \theta\, \,  \cos^2 \theta \, d \psi  \,.
\end{aligned}
\label{sol20n}
\end{equation}

There appears to be an alternative straightforward algorithm for generating solutions with $m=0$, general $n$ and larger values of $k$.
One first defines: 
\begin{equation}
\omega_{k,0,n}~=~  \hat \omega^{(\phi)}_{k,0,n} \,  \sin^2 \theta \,d\phi  ~+~ \,  \hat \omega^{(\psi)}_{k,0,n}\cos^2 \theta \, d \psi  \,.
\label{omk0n}
\end{equation}
and then makes independent Ans\"atze for $\cF_{k,0,n}$, $\hat \omega^{(\phi)}_{k,0,n}$ and $\hat \omega^{(\psi)}_{k,0,n}$ of the form:
\begin{equation}
 \sum_{j=0}^{k-1} \, F_j(r) \, \sin^{2j} \theta \,,
\label{knAnsatz}
\end{equation}
for some undetermined functions, $F_j(r)$.  As noted above in (\ref{eq:omegakmn})  and  (\ref{eq:Fkmn}), the  BPS equations for these RMS pieces of $\omega$ and $\cF$ are relatively simple.  One begins by  substituting the Ansatz for $\cF$ into (\ref{eq:Fkmn}). The result is a coupled set of ODEs involving only the functions $F_j(r)$  that, being ``upper triangular'',\footnote{That is, the equation for $F_j(r)$ only involves the $F_\ell(r)$ for $\ell \ge j$, and so one starts with the equation for $F_{k-1}(r)$ alone and then uses it to find $F_{k-2}(r)$, and in this way one continues  to the lower $F_{j}(r)$.} can iteratively solved for all the arbitrary functions.  The integration constants in these solutions are determined by requiring that solutions are regular at infinity.  Given $\cF$, (\ref{eq:omegakmn}) becomes a coupled set of first-order equations for the components of $\omega$. It is then relatively easy to cross eliminate to obtain second-order differential equations for either $\hat \omega^{(\phi)}_{k,0,n}$ or $\hat \omega^{(\psi)}_{k,0,n}$ alone. One then follows the same procedure as that used for $\cF$ to determine the  functions of $r$ and integration constants in the Ansatz (\ref{knAnsatz}).

We have implemented this procedure explicitly for $k=3$ and it generates a smooth, albeit complicated, solution that we will not present here.  

\subsection{The structure of the metric}

We now discuss the structure of the asymptotically-AdS$_3$ metrics. For concreteness we focus on the $(1,0,n)$ family of solutions.
For this family of solutions, one can prove the global absence of closed timelike curves by completing the squares on the periodic coordinates.
To display the metric in this form, following \cite{Bena:2017upb} we introduce the convenient quantity
\begin{equation}
\Lambda ~\equiv~ \frac{\sqrt{\cP}\,\Sigma}{\sqrt{Q_1 Q_5}} ~=~ \sqrt{ 1 - \frac{a^2\,b^2}{(2 a^2 +b^2)} \, \frac{r^{2n}}{(r^2 +a^2)^{n+1}} \, \sin^2 \theta  } \,.
  \label{Lambdadef1}
\end{equation}
Then the metric can be written as
\bea
ds^2 &=& - \frac{\Lambda}{\sqrt{Q_1 Q_5}} \;\! \frac{2a^2(r^2+a^2)}{2a^2+ b^2 F_0(r)} \;\! dt^2
+ \sqrt{Q_1 Q_5} \;\! \Lambda \left( \frac{dr^2}{r^2+a^2} + d\theta^2 \right) \cr
&& {} + \frac{\sqrt{Q_1 Q_5}}{\Lambda} \sin^2\theta \left( d\phi - \frac{2a^2}{2a^2+b^2} \frac{dt}{R_y} \right)^2 
\label{eq:10n-metric} \\
&& {} + \frac{R_y^2}{\sqrt{Q_1 Q_5} \;\! \Lambda} \left( a^2+\frac{b^2}{2} F_1(r) \right)  \cos^2\theta
\left( d\psi - \frac{\left(2a^2+b^2F_0(r)\right)dy + b^2 F_0(r) \;\! dt}{\left(2a^2+b^2F_1(r)\right)R_y} \right)^2 \quad~~~  \cr
&& {} + \frac{1}{\sqrt{Q_1 Q_5} \;\! \Lambda} \,
\frac{ r^2 \left( 2a^2+b^2 F_0(r) \right)  F_2(r,\theta)  }
{2a^2 \left(r^2 (2a^2+b^2) +a^2 (2a^2+b^2 F_0(r)  \right)}
\left( dy + \frac{b^2F_0(r)}{2a^2+b^2F_0(r)} dt \right)^2 \quad~~~
\nonumber
\eea
where we have used the shorthand notation
\bea
F_0(r) &=& 1-\frac{r^{2n}}{(r^2+a^2)^n} \,, \qquad F_1(r) ~=~ 1 - \frac{a^2}{r^2+a^2}\frac{r^{2n}}{(r^2+a^2)^n} \,, \cr
F_2(r,\theta) &=& r^2 (2a^2+b^2) +a^2 \left( 2a^2 +b^2 \left( 1 - \frac{r^{2n}}{(r^2+a^2)^n}\sin^2\theta \right) \right) .
\eea
We now observe that all of the angular terms have coefficients that are globally non-negative, and the only places where the coefficients vanish are at the standard degeneration of angular coordinates at $\theta = 0$ and $\theta = \pi/2$, and where the $y$ circle shrinks smoothly at $r=0$. Thus the geometry has no closed timelike curves.

To illustrate the structure of the solution, it is instructive to
examine the coefficient of $( dy +\cdots )^2$ in the last line
of~\eq{eq:10n-metric}. This controls the smooth shrinking of this fiber
at $r=0$, its stabilization at finite size in the AdS$_2$ region, and
its growth in the AdS$_3$ region.

\begin{figure}[h]
\centerline{\includegraphics[width=85mm]{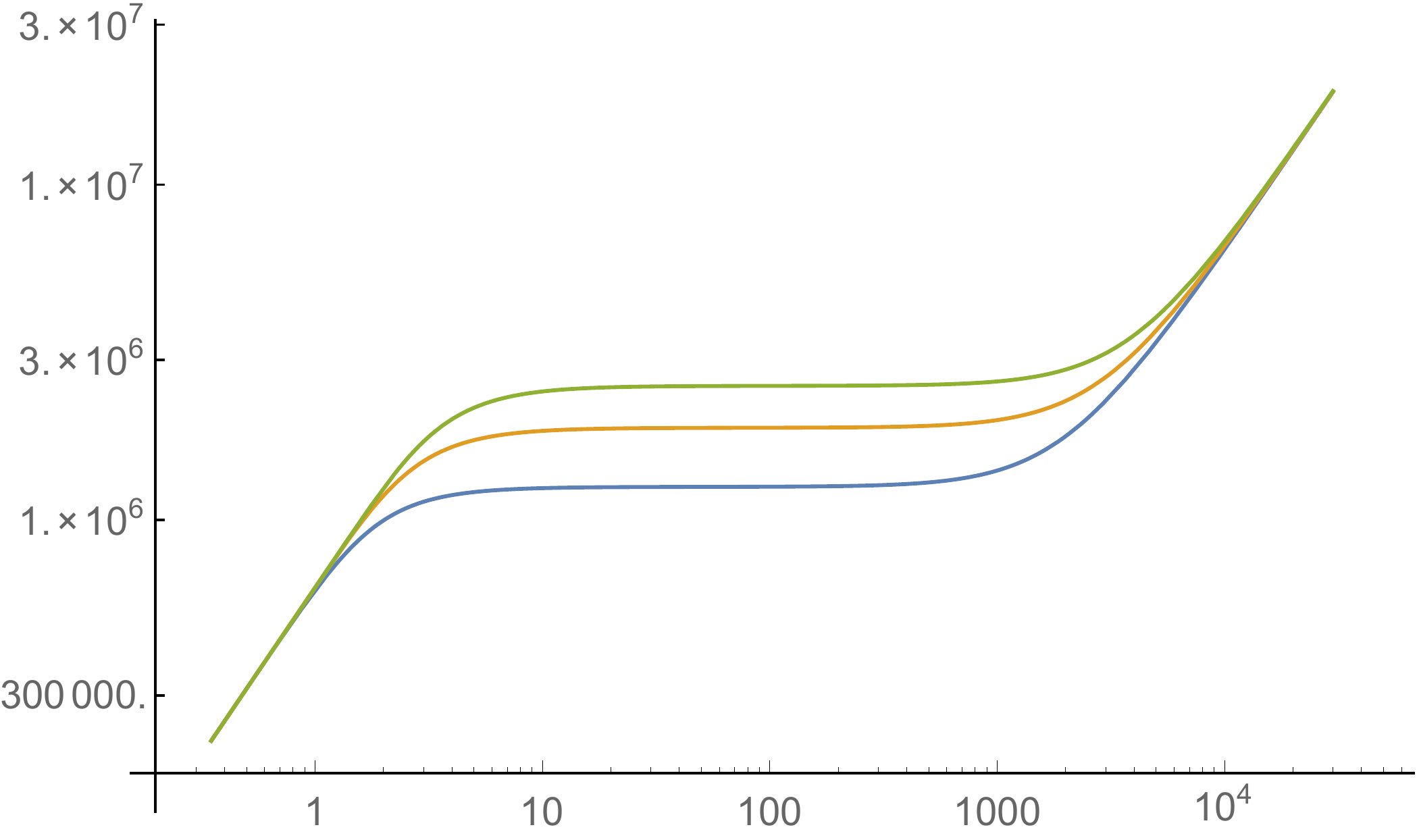}
\qquad\qquad
\includegraphics[width=57mm]{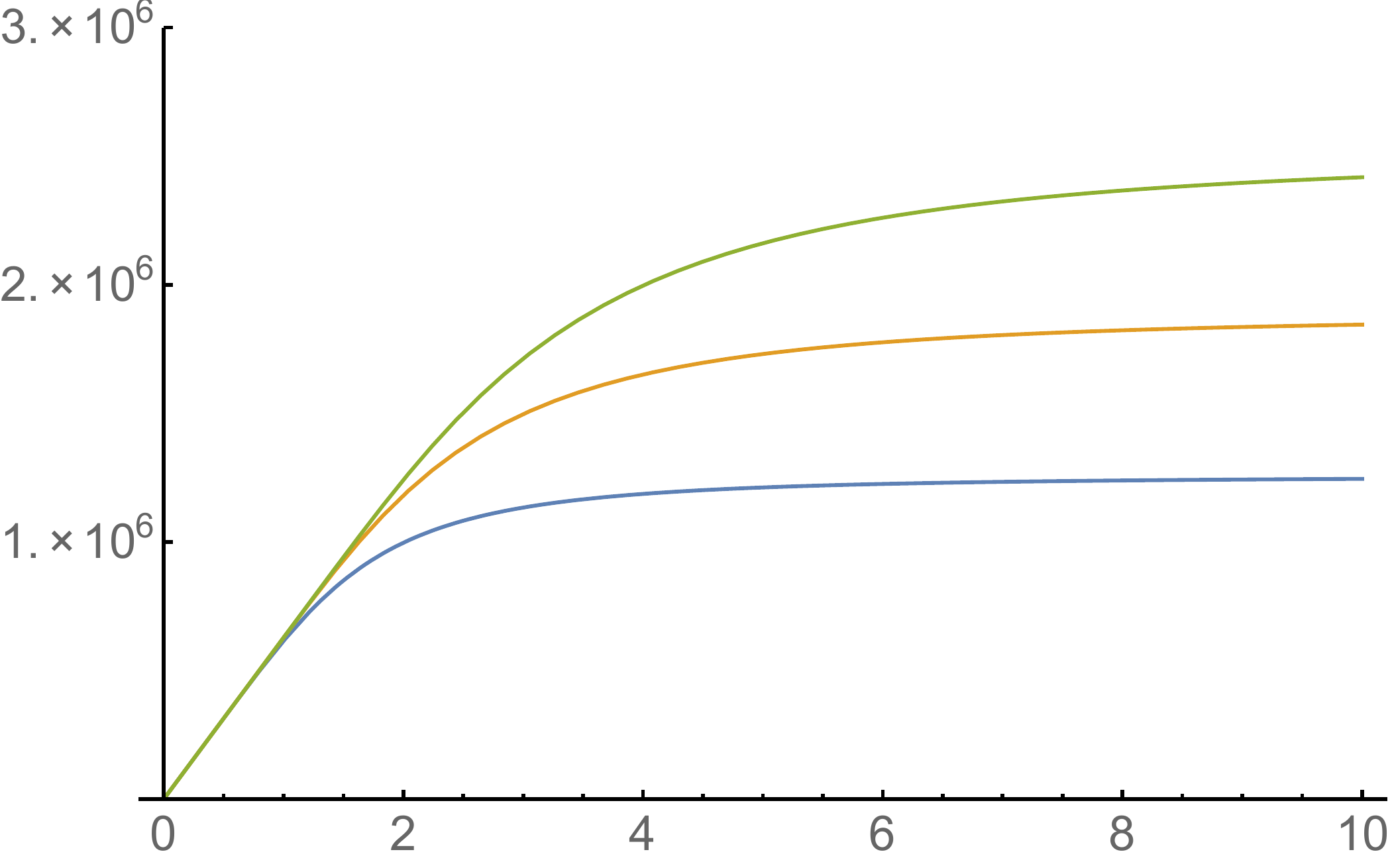}}
\caption{\it 
Left plot: Log-log plot of the proper length of a curve around the $y$ circle as described in the text (vertical axis), as a function of $r$ in units of $a$ (horizontal axis), where we set $\theta=\pi/4$. 
Right: Linear plot detailing the behaviour in the cap region. Parameters chosen: $k=1$, $m=0$, $n=\left\{ 4,9,16 \right\}$ (from bottom to top); $a=1$, $\sqrt{Q_1}=\sqrt{Q_5}=10^5$, $R_y=10^7$, so that $b \simeq \sqrt{2}\times 10^3$. 
}
\label{fig:gyy-log-log}
\end{figure}

In Figure \ref{fig:gyy-log-log} we depict this as follows. We plot the proper length of a curve where $dt=dr=d\theta=0$, where $d\phi$ and $d\psi$ are chosen to make the second and third lines of~\eq{eq:10n-metric} vanish, and where the curve traverses once around the $y$ circle. One sees the central AdS$_2$ region, where this circle has constant proper length. The regions of linear growth are the asymptotic AdS$_3$ region and the global AdS$_3$-like cap.

\section{Asymptotically-Flat Solutions}
\label{Sect:AsymFlat}

\subsection{Novel features of the asymptotically-flat solutions}

In order to construct asymptotically-flat solutions, we add $1$'s  to the warp factors  $Z_1$ and $Z_2$.  This in turn modifies the coiffuring structure and introduces extra oscillatory terms in $(Z_1,\Theta_2)$ and $(Z_4,\Theta_4)$. The first layer is simply (\ref{eq:ZIAdSsinglemode}) and (\ref{eq:ThetaIAdSsinglemode}) with the $1$'s added:   
\begin{equation}
\begin{aligned}
Z_1&~=~1\,+\,\frac{Q_1}{\Sigma} \,+\,  \frac{b_1  R_y^2 }{2 Q_5}\,\frac{\Delta_{2k,2m,2n}}{\Sigma}\, \cos(\hat{v}_{2k,2m,2n}) \,, \qquad  Z_2 ~=~ 1\,+\,\frac{Q_5}{\Sigma}  \,, \\
Z_4&~=~ b_4\,R_y\,\Delta_{k,m,n}\, \cos(\hat{v}_{k,m,n}) \,,
\end{aligned} 
\label{NewZs}
\end{equation}
with
\begin{equation}
 \Theta_1 =0\,,\qquad
 \Theta_2 = \frac{b_1 R_y}{2 Q_5}\, \vartheta_{2k,2m,2n}\,,\qquad
 \Theta_4 = b_4\, \vartheta_{k,m,n}\,.
\label{NewThetas}
\end{equation}
as in \eqref{eq:ThetaIAdSsinglemode}.

The sources for the second layer of BPS equations are now considerably more complicated: 
\begin{equation}
\begin{aligned}
& \hspace{-10mm}  Z_1 \Theta_1+ Z_2 \Theta_2   -2\,Z_4 \Theta_4 \\ 
\phantom{XXX} &~=~ \sqrt{2}\,\,R_y\, \Delta_{2k,2m,2n} \,  \left(\frac{b_4^2 -  b_1 }{\Sigma} -  \frac{b_1 }{Q_5}\right)\,\\
 & \qquad\qquad \times  \Biggl[ \left((m+n)\, r\sin\theta  +n\Bigl(  {m \over k}-1 \Bigr) \, {\Sigma\over r \sin\theta}\right)  \Omega^{(1)}\sin\vh_{2k,2m,2n}  \\
  &\hspace{28ex} +\left( m\left({ n \over k}+1\right)\Omega^{(2)}  +n\left({m\over k}-1\right)\Omega^{(3)} \right)\cos\vh_{2k,2m,2n} \Biggr] \\
 &\qquad ~+ \sqrt{2}\,R_y\, b_4^2\, \frac{\Delta_{2k,2m,2n} }{\Sigma} \,  \left( m\left({ n \over k}+1\right)\Omega^{(2)}  +n\left({m\over k}-1\right)\Omega^{(3)}\right) \,,
 \end{aligned} 
\label{source1}
\end{equation}
while the right-hand side of the second equation in (\ref{BPSlayer2}) reduces to:
\begin{equation}
\begin{aligned}
 4 \left(\frac{b_4^2 - b_1}{\Sigma} -  \frac{b_1 }{Q_5}\right)\,&\frac{(m+n)^2 \Delta_{2k,2m,2n} }{\Sigma} \,\cos\vh_{2k,2m,2n}  \\
&  ~+~ \frac{2\, b_4^2}{k^2}\, \frac{\Delta_{2k,2m,2n} }{\Sigma} \,  \left( \frac{(k-m)^2  n^2}{r^2 \sin^2 \theta}  +\frac{(k+n)^2  m^2}{(r^2 + a^2)\cos^2 \theta} \right) .
 \end{aligned} 
\label{source2}
\end{equation}

The last terms in (\ref{source1}) and  (\ref{source2}) do not depend on $(v,\phi, \psi)$ and represent the RMS effect of the modes.  These were solved in the previous section.   The new feature are the terms that depend on $\vh_{2k,2m,2n}$, that have coefficient:  
\begin{equation}
\left(\frac{b_4^2 - b_1}{\Sigma} -  \frac{b_1 }{Q_5}\right) \,.
\label{osccoeff}
\end{equation}
The  constant term proportional to $ b_1 $ term is a new contribution, coming from the $1$'s in the $Z_I$'s, while the $(b_4^2 - b_1)$ term was removed earlier by coiffuring.  
Because of the explicit $(r,\theta)$-dependence of the complete coefficient, (\ref{osccoeff}),  the oscillating modes cannot be completely removed via coiffuring.  The second layer of BPS equations must therefore be solved directly, with all the sources, and a modified coiffuring condition will then be determined by removing singularities from the complete solution.

\subsection{The second layer of equations}

We now focus entirely on the oscillating parts of (\ref{source1}) and  (\ref{source2}).    These are consistent with the Ansatz:
\begin{align}
\begin{split}
 \omega^{\rm osc} &= (\hat \omega_r \, dr +   \hat\omega_\theta \, d\theta)\,\sin\vh_{2k,2m,2n}
 +   (\hat \omega_1 \, d\phi +   \hat\omega_2 \, d\psi)\,\cos\vh_{2k,2m,2n} \,,\\
  \cF^{\rm osc}  &= \widehat F\,  \cos\vh_{2k,2m,2n} \,.
\end{split}
\label{Fomansatz}
\end{align}
One then decomposes this equation into differential operators 
\begin{equation}
\begin{aligned}
\mathcal{D} \omega^{\rm osc} + *_4  \mathcal{D}\omega^{\rm osc}  + \mathcal{F}^{\rm osc}\,d\beta ~=~  &(r^2 +a^2)\, \cos\theta \,\sin\vh_{2k,2m,2n} \,\Omega^{(1)}\, \cL_1^{(2k,2m,2n)} \\ & 
+ \frac{1}{r}(r^2 +a^2)\, \cos\theta \,\cos\vh_{2k,2m,2n} \,\Omega^{(2)} \, \cL_2^{(2k,2m,2n)}  \\ &
+ r\, \sin\theta \,\cos\vh_{2k,2m,2n} \,  \Omega^{(3)} \,\cL_3^{(2k,2m,2n)} \,, \\
 *_4\mathcal{D} *_4\!\big(\dot{\omega}^{\rm osc} - \coeff{1}{2}\,\mathcal{D} \mathcal{F}^{\rm osc} \big)  ~=~&  \coeff{1}{2}\,\cos\vh_{2k,2m,2n} \, \cL_4^{(2k,2m,2n)}  \,, 
 \end{aligned}
 \end{equation}
 where
 \begin{equation}
\begin{aligned}
\mathcal{L}^{(2k,2m,2n)}_1 ~=~ & (\partial_r\hat \omega_\theta-\partial_\theta\hat \omega_r) -\frac{2}{r\, (r^2+a^2)\,\sin\theta\,\cos\theta}\,\big[\big((m+n) r^2 - n \Sigma\big) \hat \omega_1 \\& 
\qquad\qquad\qquad\qquad\qquad\qquad\qquad\qquad + \big ((k+n) \Sigma -(m+n)(r^2+a^2)  \big)\,\hat \omega_2 \big]\,,\\
\mathcal{L}^{(2k,2m,2n)}_2 ~=~ &\frac{1}{\cos\theta}\,\partial_r\hat \omega_2  + \frac{r}{(r^2+a^2) \,\sin\theta}\, \partial_\theta \hat \omega_1 \\ 
& - \frac{2}{\Sigma \,\sin\theta\,\cos\theta }\,\Big[  \frac{r\,\cos\theta}{(r^2+a^2)}\,\big((k+n)\Sigma - (m+n)(r^2+a^2) \big) \,\hat \omega_\theta   \\ &
\qquad \qquad\qquad\qquad - \sin\theta \,\big((m+n) r^2  - n \Sigma\big) \,  \hat \omega_r \Big] ~+~  \frac{\sqrt{2}\,R_y\, a^2  \, r \cos\theta }{\Sigma^2} \,  \widehat F\,,\\
\mathcal{L}^{(2k,2m,2n)}_3 ~=~ &\frac{1}{\sin\theta}\,\partial_r\hat \omega_1  - \frac{1}{r \,\cos\theta}\, \partial_\theta \hat \omega_2 \\ 
& - \frac{2}{\Sigma \,r\,\sin\theta\,\cos\theta }\,\Big[r \cos\theta \,\big((k+n)\Sigma - (m+n)(r^2+a^2) \big) \,\hat \omega_r   \\ &
\qquad \qquad\qquad\qquad + \sin\theta \,\big((m+n) r^2  - n \Sigma\big) \,  \hat \omega_\theta \Big] ~-~  \frac{\sqrt{2}\,R_y\, a^2  \, r \sin\theta }{\Sigma^2} \,\widehat F\,, \\
\mathcal{L}^{(2k,2m,2n)}_4  ~=~   & \cL_0^{(2k,2m,2n)}  \widehat F~-~ \frac{4\sqrt{2}\,(m+n)}{R_y}\,{\rm div}^{(2k,2m,2n)} \hat \omega \,,
\end{aligned}
\label{L1to3defns}
\end{equation}
and
\begin{equation}
\begin{aligned}
\mathcal{L}_0^{(2k,2m,2n)} \widehat F  ~\equiv~& \frac{1}{\Sigma}\, \bigg[\frac{1}{r}\,\partial_r \big(r (r^2+a^2) \partial_r \widehat F\big) ~+~ \frac{1}{\sin\theta\,\cos\theta}\,\partial_\theta \big(\sin\theta\,\cos\theta \, \partial_\theta \widehat F\big) \\
& \qquad ~-~4\, \bigg(\frac{n^2\, a^2}{r^2} -\frac{(k+n)^2\, a^2}{(r^2+a^2)}+\frac{(k-m)^2}{\sin^2\theta } + \frac{m^2}{\cos^2\theta } \bigg)\,  \widehat F\, \bigg]   \,,\\
{\rm div}^{(2k,2m,2n)} \omega^{\rm osc} ~\equiv~   & \frac{1}{\Sigma}\, \bigg[\frac{1}{r}\,\partial_r \big(r (r^2+a^2) \hat \omega_r \big) ~+~ \frac{1}{\sin\theta\,\cos\theta}\,\partial_\theta \big(\sin\theta\,\cos\theta \,  \hat \omega_\theta \big) \\
& \qquad- \frac{2}{(r^2+a^2) \,\sin^2\theta}\, \big((k+n)\Sigma - (m+n) (r^2+a^2)\big) \,\hat \omega_1\\
& \qquad + \frac{2}{r^2\,\cos^2\theta}\, \big((m+n)r^2 - n \Sigma\big) \,\hat \omega_2\bigg] \,.   
\end{aligned}
\label{L0tand4defns}
\end{equation}
Using the sources (\ref{source1}) and (\ref{source2}) we arrive at the following equations:
 \begin{equation}
\begin{aligned}
\mathcal{L}^{(2k,2m,2n)}_1 ~=~ &  \frac{ \sqrt{2}\,\,R_y\,  \Delta_{2k,2m,2n}}{(r^2+a^2) \cos\theta} \,  \left(\frac{b_4^2 - b_1}{\Sigma} -  \frac{b_1 }{Q_5}\right)\,\left((m+n)\, r\sin\theta  +n\Bigl(  {m \over k}-1 \Bigr) \, {\Sigma\over r \sin\theta}\right)  , \\
\mathcal{L}^{(2k,2m,2n)}_2 ~=~&  \frac{\sqrt{2}\,\,R_y\,r\,  \Delta_{2k,2m,2n}}{(r^2+a^2) \cos\theta} \,  \left(\frac{b_4^2 - b_1}{\Sigma} -  \frac{b_1 }{ Q_5}\right)\,   m\left({ n \over k}+1\right) ,   \\ 
\mathcal{L}^{(2k,2m,2n)}_3 ~=~ & \frac{ \sqrt{2}\,\,R_y\,  \Delta_{2k,2m,2n}}{r \sin\theta} \,  \left(\frac{b_4^2 -b_1}{\Sigma} -  \frac{b_1 }{Q_5}\right)\,n\left({m\over k}-1\right) ,\\ 
\mathcal{L}^{(2k,2m,2n)}_4  ~=~&   \frac{8\, (m+n)^2\,  \Delta_{2k,2m,2n}}{\Sigma} \,    \left(\frac{b_4^2 - b_1}{\Sigma} -  \frac{b_1 }{Q_5}\right) .
\end{aligned}
\label{Layer2eqns}
\end{equation}

Given that the BPS solution is $u$ independent, any BPS solution is invariant under the following reparametrization of $u$:
\begin{equation}\label{gaugecoords}
u\to u + U(x^i,v)\,,\quad \omega \to  \omega - dU + \dot{U}\,\beta \,,\quad \mathcal{F} \to \mathcal{F} - 2\, \dot{U}\,,
\end{equation}
This leads to the gauge invariance: 
\begin{equation}\label{gaugetrf}
\begin{aligned}
  \omega^{\rm osc}  ~\to~ &   \omega^{\rm osc}  + (\partial_r f \, dr+ \partial_\theta f \, d\theta)\,\sin\vh_{2k,2m,2n}  \\
& + \frac{f}{\Sigma} \, \Big[ \big((k-m) (r^2+a^2) - a^2  (k+n)\sin^2\theta \big)\, d \phi -  \big(m r^2  - n a^2 \cos^2\theta \big)\, d \psi  \Big]\,\cos\vh_{2k,2m,2n} \,,\\
\widehat F ~\to~ & \widehat F ~+~  \frac{2\sqrt{2}\,(m+n)}{R_y}\, f \,,
 \end{aligned}
\end{equation}
for any function, $f(r,\theta)$. For our oscillating modes we will use this gauge invariance to set: 
\begin{equation}
\widehat F ~=~ \cF^{\rm osc} ~=~  0\,.
\label{gaugechoice}
\end{equation}
%

\subsection{Solving the second layer}
\label{Solving-layer2}

The standard route to solving the system (\ref{Layer2eqns}) is to observe that the equations involving $\mathcal{L}^{(2k,2m,2n)}_2$ and $\mathcal{L}^{(2k,2m,2n)}_3$ do not involve derivatives of $\hat \omega_r$ and $\hat \omega_\theta$.  One then uses these equations to obtain expressions for $\hat \omega_r$ and $\hat \omega_\theta$ and then substitutes them back into the other two equations to obtain two second order differential equations for $\hat \omega_1$ and $\hat \omega_2$.  Rather remarkably, one then finds that the combination $\hat \omega_1+\hat \omega_2$ satisfies a straightforward harmonic equation involving $\mathcal{L}^{(2k,2m,2n)}_0$. 

From the equations above, we find:
\begin{equation}
\begin{aligned}
& \mathcal{L}^{(2k,2m,2n)}_0 (\hat \omega_1 + \hat \omega_2)  \\
& \quad ~=~2\sqrt{2}\,R_y\, \Delta_{2k,2m,2n} \, \left(\frac{a^2 \, (m+n)(b_4^2 -   b_1)}{\Sigma^2} -  \frac{b_1 }{k\, Q_5}\, (m(k+n) -n(k-m)) \,\right)  .
\end{aligned}
\label{omsumeqn1}
\end{equation}
It is elementary to solve this and we find the following particular solution: 
\begin{equation}
 (\hat \omega_1 + \hat \omega_2)   ~=~ - \frac{R_y}{2\sqrt{2}} \, \Delta_{2k,2m,2n} \, \left(\frac{(b_4^2 - b_1)}{\Sigma} -   \frac{b_1 }{k^2\, Q_5}\, (m(k+n) -n(k-m)) \,\right)  .
\label{omsumsol1}
\end{equation}

The next step is slightly more of an art than a science.  The individual equations for  $\hat \omega_1$ and $\hat \omega_2$ separately are very complicated.  However, based on experience, the form of $\hat \omega_1 + \hat \omega_2$, and how $\hat \omega_1$ and $\hat \omega_2$ should behave in various limits, one is naturally led to 
\begin{equation}
\begin{aligned}
\hat \omega_1   ~=~ & \frac{R_y}{2\sqrt{2}} \, \Delta_{2k,2m,2n} \, \left(-(b_4^2 -b_1)\, \frac{(r^2+a^2)}{a^2\, \Sigma} +   \frac{b_1 }{Q_5}\,\frac{m(k+n)}{k^2}\,\right)  , \\ 
\hat \omega_2   ~=~&  \frac{R_y}{2\sqrt{2}} \, \Delta_{2k,2m,2n} \, \left((b_4^2 - b_1) \,\frac{r^2}{a^2\,\Sigma} -  \frac{b_1 }{Q_5}\,\frac{n(k-m)}{k^2}\,\right)  .
\end{aligned}
\label{omsol1}
\end{equation}
These manifestly add to (\ref{omsumsol1}), however these expressions are \emph{not the solutions for general} $(k,m,n)$ but they \emph{are solutions for either $m=0$ or $m=k$}.  Thus we will preserve the appearance of $m$ in our formulae with the understanding, for the moment, that we are considering $m=0$ or $m=k$.  Presumably there are more complicated recurrence relations for  solutions with intermediate values of $m$.

Armed with expressions for  $\hat \omega_1$ and $\hat \omega_2$, one can now substitute back into the equations in (\ref{Layer2eqns}) involving $\mathcal{L}^{(2k,2m,2n)}_2$ and $\mathcal{L}^{(2k,2m,2n)}_3$  and solve for $\hat \omega_r$ and $\hat \omega_\theta$ algebraically.  The general result is a mess, but there are simple formulae that work for $m=0,k$:
\begin{equation}
\begin{aligned}
\hat \omega_r   ~=~ & -  \frac{b_1 \,R_y}{2 \sqrt{2} \, Q_5}\, \Delta_{2k,2m,2n} \, \frac{k(m+n)r^2 + n (k-m) a^2}{k^2 \, r (r^2 + a^2)}  \,, \\ 
\hat \omega_\theta   ~=~&  \frac{R_y}{2\sqrt{2}} \, \frac{\Delta_{2k,2m,2n} }{k^2\, a^2\, \sin\theta \, \cos\theta}\, \Big(k(2m-k) (b_4^2 - b_1) \\
& \qquad\qquad\qquad\qquad\qquad  + \frac{b_1 \, a^2}{Q_5}\,\big((m+n) ((k-m) \sin^2 \theta-m \cos^2 \theta\big)+m (k-m)\Big) \,.
\end{aligned}
\label{omsol2}
\end{equation}
So far (\ref{omsol1}),  (\ref{omsol2}) and (\ref{gaugechoice})  define complete solutions for $\omega^{\rm osc}$ for $m=0$ and $m=k$.

The careful reader might note that we have added a seemingly redundant $m(k-m)$ term to the expression for $\hat \omega_\theta$.  This is because if one substitutes (\ref{omsol1}),  (\ref{omsol2}) and (\ref{gaugechoice}) into (\ref{Layer2eqns}) then the result either vanishes or is proportional to 
\begin{equation}
m(k-m)\bigg((b_4^2 - b_1) -   \frac{b_1 \, a^2}{Q_5}\, \frac{(m+n)}{k}\bigg)\,.
\label{coiff1}
\end{equation}
Thus  (\ref{omsol1}),  (\ref{omsol2}) and (\ref{gaugechoice}) provides a solution \emph{for all} $(k,m,n)$ provided that 
\begin{equation}
\bigg((b_4^2 - b_1) -   \frac{b_1 \, a^2 }{Q_5}\, \frac{(m+n)}{k}\bigg)  ~=~ 0\,.
\label{coiff2}
\end{equation}
As we will see below, this is the new coiffuring constraint required by regularity of the solution and so we actually have the \emph{complete, regular solution  for all} $(k,m,n)\,${\em !}

The way we first arrived at this complete solution was to find the coiffuring constraint for $m=0$ and $m=k$, and from this we inferred the general coiffuring relation (\ref{coiff2}).  Then we used $\hat \omega_1$ and $\hat \omega_2$ in (\ref{Layer2eqns}) to solve for $\hat \omega_r$ and $\hat \omega_\theta$ algebraically and then imposed  (\ref{coiff2}).  This  led to the complete expressions for $\hat \omega_r$ and $\hat \omega_\theta$.  The complete solution for $\omega^{\rm osc}$  is given by (\ref{Fomansatz}) with components given by (\ref{omsol1}) and (\ref{omsol2}).

Putting the components together and using the coiffuring  constraint (\ref{coiff2}), we can simplify  $\omega^{\mathrm{osc}}$ to: 
\bea
\omega^{\mathrm{osc}} &=& 
-\frac{b_1}{Q_5}\frac{R_y}{2 \sqrt{2}}\Delta_{2k,2m,2n}\Bigg\{
\left(\frac{(m+n)}{k}\frac{a^2 \sin^2\theta}{\Sigma}+\frac{n(k-m)}{k^2}\right)
\cos{\hat{v}_{2k,2m,2n}} \;\! d\phi \cr
&&{}\qquad\quad + \left( \frac{(m+n)}{k}\frac{a^2 \cos^2\theta}{\Sigma}-\frac{m(k+n)}{k^2} \right) \cos{\hat{v}_{2k,2m,2n}} \;\! d\psi \cr
&&{}\qquad\quad + \left( r^2 \frac{(m+n)}{k} +a^2 \frac{n(k-m)}{k^2} \right) 
\frac{1}{r(r^2+a^2)} \sin{\hat{v}_{2k,2m,2n}} \;\! dr \label{eq:omegaext}\\
&&{}\qquad\quad + \left(  \frac{n(k-m)}{k^2} \cot\theta - \frac{m(k+n)}{k^2} \tan\theta \right)  \sin{\hat{v}_{2k,2m,2n}} \;\! d\theta\Bigg\} \,. 
\nonumber
\eea

Finally, we note that  the coiffuring condition may be re-written as:
\bea
b_1 \left( 1+ \frac{a^2}{Q_5}\frac{m+n}{k} \right) = b_4^2 \,.
\eea
This form is useful because $a^2/Q_5$ is a small dimensionless parameter in the near-decoupling limit.

\subsection{Asymptotically-Flat Solutions: Regularity and Conserved Charges}

The complete asymptotically-flat solution to the second layer of the BPS equations is given by: 
\bea \label{eq:omega-parts}
\omega = \omega_0 + \omega^{\mathrm{RMS}}+\omega^{\mathrm{osc}} \,, \qquad \cF ~=~ \cF^{\mathrm{RMS}} \,,
\eea
where the individual pieces are given by (\ref{seed_O(b)_omega_F}), (\ref{eq:omkmnparts1}) and (\ref{eq:omegaext}).


The general conditions for regularity have been discussed in Section \ref{Sect:regularityAdS}. We verify here that these conditions are satisfied also by the asymptotically-flat extension of our solutions. We focus on the two potentially problematic points: the center of $\mathbb{R}^4$ $(r=0, \theta=0)$ and the supertube location $(r=0, \theta=\pi/2)$.

\subsubsection{Near $(r=0, \theta=0)$}

In the asymptotically-flat solution the one-form $\omega$ acquires a new contribution $\omega^{\mathrm{osc}}$ that depends on $\hat v_{2k,2m,2n}$, and the analysis of its behaviour around the point $(r=0, \theta=0)$, where the polar coordinates degenerate, requires some extra care. Notice first of all that $\omega^{\mathrm{osc}}$ is finite at $(r=0, \theta=0)$, since the $1/r$ and $1/\sin\theta$ poles inside the curly bracket in (\ref{eq:omegaext}) are canceled by $\Delta_{2k,2m,2n}$. This is not enough however to conclude that $\omega^{\mathrm{osc}}$ is smooth: we should check that its components with respect to a local orthonormal frame are finite. Switching to the coordinates $(\tilde r,\tilde \theta)$ and $\tilde v$ defined in (\ref{eq:tildertheta}) and (\ref{eq:vtilde}), we find
\bea
\omega^{\mathrm{osc}}&\sim& \frac{n(k-m)}{k^2}\,\Delta_{2k,2m,2n} \,\left[2\,\sin\hat v_{2k,2m,2n} \left(\frac{d\tilde r}{\tilde r}+\cot 2\tilde \theta\, d\tilde \theta\right) + \cos\hat v_{2k,2m,2n}\, (d\phi+d\psi) \right]\nonumber\\
&\sim& \frac{1}{k^2}\,\mathrm{Im}\bigg [e^{\frac{i(m+n)\,2\sqrt{2}\,\tilde v}{R_y}}\Big ((k-m) (\tilde r \sin\tilde \theta e^{i\phi})^{2(k-m)}\, d (\tilde r \cos\tilde\theta e^{i\psi})^{2n}\nonumber\\
&&\qquad\quad+n (\tilde r \cos\tilde\theta e^{i\psi})^{2n} \,d (\tilde r \sin\tilde \theta e^{i\phi})^{2(k-m)} \Big)\bigg]\,.
\eea
Since $\tilde r \sin\tilde \theta e^{i\phi}$ and $\tilde r \cos\tilde\theta e^{i\psi}$ are linear combinations of well-behaved Cartesian coordinates around $(r=0, \theta=0)$, the identity above shows that $\omega^{\mathrm{osc}}$ is smooth at the center of space.

\subsubsection{Near $(r=0, \theta=\pi/2)$}

Near $r=0, \theta=\pi/2$, one can make the coordinate transformation
\bea
r = a \lambda \cos\chi \,, \qquad \theta = \frac{\pi}{2}-\lambda\sin\chi
\eea
where we consider $\lambda$ to be a small parameter. 

Recalling that near $r=0, \theta=\pi/2$, $\Delta_{2k,2m,2n}$ behaves like
\bea
\Delta_{2k,2m,2n} &\sim & r^{2n} (\cos\theta)^{2m} \,,
\eea
and noting that we always have at least one of $n$ or $m$ greater than zero, 
we see that in the above coiffured $\omega^{\mathrm{osc}}$ there are no terms that scale as $\lambda^{-1}$  when $\lambda\to 0$ and that the first terms start at $\lambda^0$.
Thus, near $\Sigma=0$, $\omega$ is well-approximated by $\omega^{\mathrm{RMS}}$.

Therefore the requirement that the $1/\Sigma$ terms in the metric near $\Sigma=0$ vanish is the same as in the asymptotically-AdS solutions, and leads to the constraint (\ref{eq:Q1Q5}). Having ensured this, the solution is smooth in the neighborhood of $\Sigma=0$.

\subsection{Conserved charges}

The global charges are read off from the asymptotically-flat solution in
a straightforward way. The oscillating terms average to zero when
integrated over the $\mathbb{S}^1$ and hence give vanishing
contributions to the global charges. Only the RMS modes, which were
derived in Section \ref{Sect:AdS}, are therefore relevant for this
computation. Moreover, since the interaction between different modes
produces terms with a non-trivial $v$-dependence which also do not
contribute to the charges, the relations valid for general multi-mode
solutions are given by simply summing the contributions of the single
modes that we write below.

The D1 and D5 supergravity charges $Q_1$ and $Q_5$ are given by the $1/r^2$ terms in the large $r$ expansion of the warp factors $Z_1$ and $Z_2$. As was noted before, regularity imposes the constraint (\ref{eq:Q1Q5}) on $Q_1$ and $Q_5$. The dimensionful momentum charge $Q_p$ is likewise encoded in the function $\mathcal{F}$ as $\mathcal{F}\approx -2 Q_p/r^2$. The expansion of  (\ref{cF}) gives
\begin{equation}
\label{eq:Qpsugra}
Q_p = \bigl(b_4^{k,m,n}\bigr)^2 \;\frac{m+n}{2 k}\, {k \choose m}^{-1} {k+n-1 \choose n}^{-1}\,.
\end{equation}
The dimensionful angular momenta $J$, $\tilde J$ can be extracted from the $\psi+\phi$ component of the one-form $\beta+\omega$:
\begin{equation}
\label{beta_omega_and_J}
\beta_\psi+\omega_\psi + \beta_\phi+\omega_\phi\approx \sqrt{2}\,\frac{J- \tilde J \cos2\theta }{r^2}\,,
\end{equation}
of which we know a closed form expression for any $k,m,n$, given in (\ref{eq:beta}), (\ref{seed_O(b)_omega_F}), (\ref{mu}). One finds
\begin{equation}
\label{eq:Jsugra}
J = \frac{R_y}{2}\left[ a^2 + \bigl(b_4^{k,m,n}\bigr)^2\; \frac{m}{k} \, {k \choose m}^{-1} {k+n-1 \choose n}^{-1}\right]\,,\quad \tilde J = \frac{R_y}{2}\, a^2\,.
\end{equation}

One can check that the charges computed from the asymptotically-flat
solution are identical to those obtained from the AdS geometry. These
can be compared with the charges of the dual CFT states.  In Section
\ref{Sect:CFT}, we will see that the supergravity and CFT charges agree 
if we assume simple linear relations between the amplitude
parameters in supergravity, $a,b_4^{k,m,n}$, and the corresponding
parameters in CFT\@.

The most significant feature of our solutions is that they can be taken to lie deep within the black hole regime $n_1 n_5 n_p-j^2>0$, {\it i.e.}~the regime of parameter space where black holes with a regular horizon exist.  We observe that our solutions lie within this bound for
\begin{equation}
  \label{eq:bhbound}
  \frac{b^2}{a^2}> \frac{k}{n+\sqrt{(k-m+n)(m+n)}}\,.
\end{equation}

\section{CFT states dual to the Asymptotically-AdS solutions}
\label{Sect:CFT}

The geometries we have constructed have macroscopic brane charges. 
As is usual in gauge/gravity duality, one can go to a region
of the moduli space where the geometry near the branes decouples
from the ambient spacetime, and correspondingly the dynamics on the
branes decouples from gravity in the asymptotically-flat region.  Quantum gravity
in the near-source geometry is then dual to a non-gravitational
theory~\cite{Maldacena:1997re}.  The asymptotically-\adsthree\  solutions of
Section~\ref{Sect:AdS} are dual to states in the 2d CFT that arises as
the low-energy limit of the gauge theory on the underlying system of branes.
In the next subsection we review some basic properties of this CFT and in 
Sections \ref{Sect:dual-CFT-states} and \ref{Sect:comparison} we 
identify the CFT states dual to the geometries we construct.

\subsection{The CFT moduli space and the symmetric orbifold}

In the weak-coupling limit of the dynamics of $n_5$ D5-branes, the $n_1$ D1-branes bind to the D5 branes by dissolving into them as instanton strings%
~\cite{Vafa:1995bm,Douglas:1995bn,Maldacena:1997re}. 
The corresponding CFT is thus often thought of as a sigma model on the moduli space of $n_1$ instantons in $U(n_5)$ gauge theory on $\cM=\bbT^4$ or $K3$.  This description of the CFT is however an approximation adapted to a particular corner of the CFT moduli space.
Consider for instance supergravity compactified on $\IT^4\times\IS^1_y$; it has a moduli space 
\be
\Bigl(\frac{E_{6(6)}}{\it USp(8)}\Bigr)\Big/ E_{6(6)}(\IZ) ~.
\ee
The decoupling limit takes $R_y/\lstr\to\infty$ holding the energy scale $E R_y$ and the $\IT^4$ volume $v_4\equiv V_4/\lstr^4$ fixed; one is effectively going to the cusp in the moduli space where $R_y$ is asymptotically large, and in particular $\sqrt{Q_1 Q_5}\ll R_y$.  In the geometry sourced by the branes, the limit isolates the region $r^2\ll Q_1, Q_5$.

The decoupling limit breaks the duality symmetry to $SO(5,5;\IZ)$, and the remaining moduli in the cusp parametrize the space
\be
\label{T4moduli}
\Bigl(\frac{SO(5,5)}{SO(5)\times SO(5)}\Bigr)\Big/ SO(5,5;\IZ) ~.
\ee
The $\underline{\bf 27}$ of wrapped brane and momentum charges on $\IT^4\times\IS^1$ splits up into 
${\underline{\bf 10}}\oplus{\underline{\bf 16}}\oplus{\underline{\bf 1}}$, where the ${\underline{\bf 10}}$ consists of branes wrapping $\IS^1$ which become infinitely heavy in the decoupling limit, and thus are part of the background data of the CFT;
the ${\underline{\bf 16}}$ consists of the assortment of branes wrapping $\IT^4$ but not $\IS^1$; and the ${\underline{\bf 1}}$ is the momentum charge on $\IS^1$.
The background of $n_5$ D5-branes and $n_1$ D1-branes breaks the duality symmetry further; of the 25 moduli in~\eqref{T4moduli}, five are frozen by the attractor mechanism~\cite{Ferrara:1995ih,Ferrara:1996dd}, and the duality group is broken to the subgroup 
$\cH_\Gamma $ of the duality ``little group'' $SO(5,4;\IZ)$ 
which fixes the ten-component background charge vector $\Gamma$.  Similar considerations hold for $\cM=K3$.  In the end, for $\cM=\IT^4$ the CFT has a 20-dimensional moduli space of couplings
\be
\cM_\cX^{\strut} = \Bigl(\frac{SO(5,4)}{SO(5)\times SO(4)}\Bigr)\Big/\cH_\Gamma ~.
\ee
The structure is conveniently seen by isolating an $SO(2,2;\IZ) = SL(2,\IZ)_L\times SL(2,\IZ)_R$ subgroup of the modular group that acts on the moduli $\tau = C_0+ i/\gstr$ by $g_R$ fractional linear transformations and $\tilde\tau = C_4+iv_4/\gstr$ by $g_L$ fractional linear transformations (when all the other antisymmetric tensor moduli are set to zero).  The background charges $(\nt_1,n_1,\nt_5,n_5)$ of fundamental and D-strings, NS5 and D5-branes, respectively, can be packaged into a matrix
\be
Q =  \begin{pmatrix} \nt_1 & n_1\\ -n_5 & \nt_5 \end{pmatrix} 
\ee
which transforms under duality as $Q\to g_L Q g_R^t$ and in particular preserves $N=\det(Q)$; we are interested in the duality frames where $\nt_1=\nt_5=0$ and $n_1n_5=N$.  The attractor mechanism then relates $\tau$ and $\tilde \tau$ via $\tilde\tau = \tau d_1/d_5$.

\begin{figure}[h]
\centerline{\includegraphics[width=3.3in]{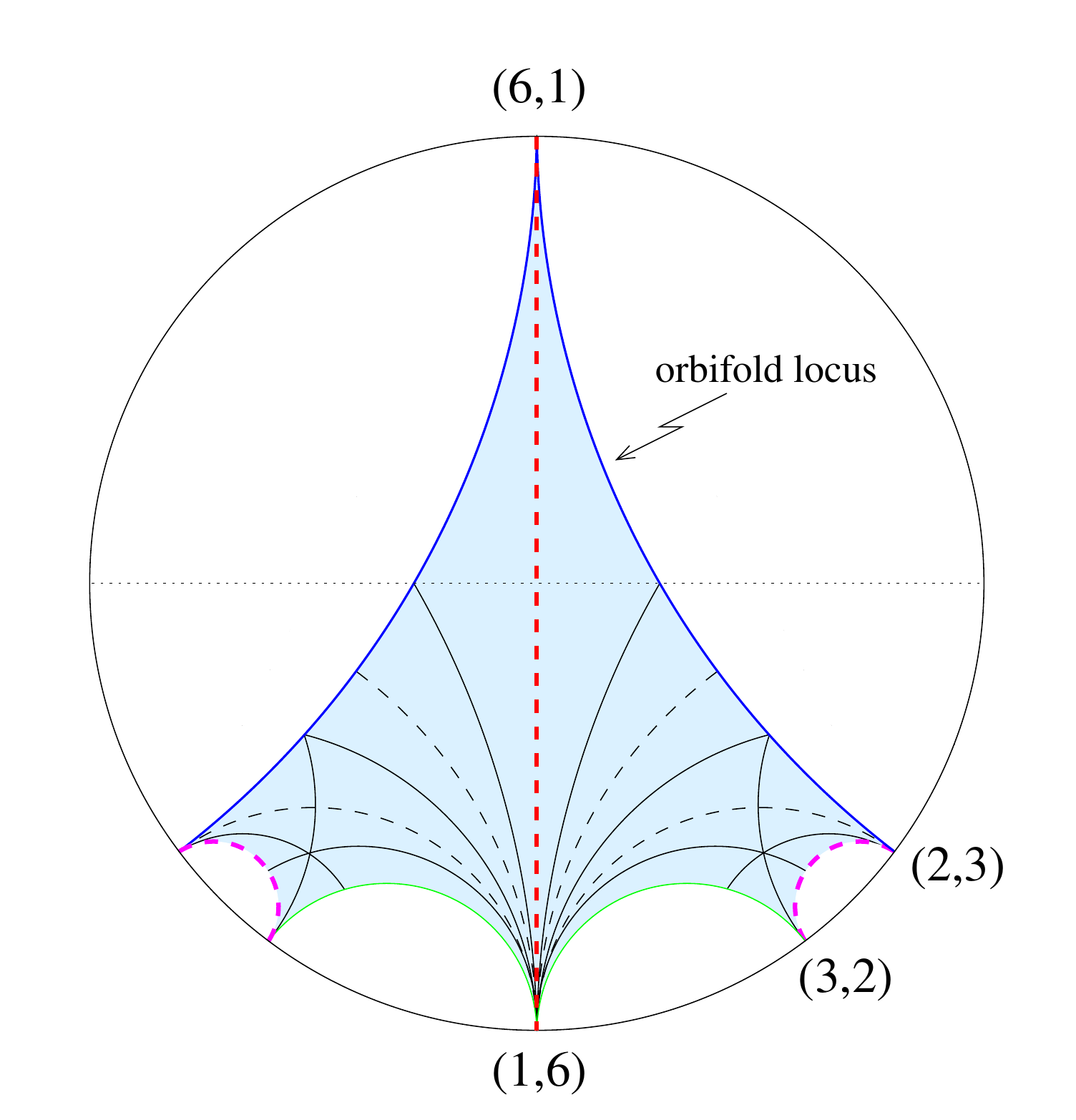}}
\setlength{\unitlength}{0.1\columnwidth}
\caption{\it 
The moduli space of the coupling $\tau=C_0+i/\gstr$ (here mapped from the upper half-plane to the Poincar\'e disk) has a cusp for every decomposition of $N$ into two factors ($n_1,n_5$) such that $N=n_1n_5$.  This slice of the moduli space is the fundamental domain of the congruence subgroup $\Gamma_0(N)$ of $SL(2,\IZ)$; $n_5$ copies of the $SL(2,\IZ)$ fundamental domain meet at the cusp corresponding to backgrounds with $n_5$ fivebranes.  Here we illustrate the structure for $N=6$.  
}
\label{fig:cuspform}
\end{figure}

The moduli space has a cusp for every factorization of the integer $N$ into a pair of integers $n_1$ and $n_5$~\cite{Seiberg:1999xz,Larsen:1999uk}, see Figure~\ref{fig:cuspform}.%
\footnote{We assume that in the prime factorization of $N$, no prime occurs more than once, so that in every cusp, $n_1$ and $n_5$ are coprime, so that the brane background is truly bound and cannot fragment into smaller pieces.}  
One sees this from the duality rotation with
\be
\label{dualityrotation}
g_L = \begin{pmatrix} a n_5& b n_1\\ 1 & 1 \end{pmatrix}
~~,~~~~
g_R = \begin{pmatrix} a & b \\ n_1 & n_5 \end{pmatrix}
\ee
which maps the D1-D5 charges $(n_1,n_5)$ to $(n_1n_5,1)$, and relates a cusp at $\tau=\frac{a}{n_1}$ to the cusp at $\tau=i\infty$, and a cusp at $\tau=\frac{b}{n_5}$ to the cusp at $\tau=0$.  Note that these two cusps are always separated by $\frac{a}{n_1}-\frac{b}{n_5}=\frac1{n_1n_5}$.
%
While this is \emph{not} a duality transformation that preserves the background, the fact that the moduli space is a symmetric space under the action of continuous duality rotations in $SO(5,4)$ means that if there is a cusp for a particular choice of charges $(n_1,n_5)$, then there is another cusp with the charges $(n_1n_5,1)$, or for that matter any pair of integers whose product is $N$.  To get from one to the other involves moving a macroscopic distance through the moduli space from one cusp to another.

In each cusp, there is a codimension-four singular locus where the system is neutrally stable and can fragment by breaking apart into separate charge centers~\cite{Seiberg:1999xz,Larsen:1999uk}.  For instance, the long string sector of perturbative string theory in ${\rm AdS}_3\times \mathbb{S}^3$~\cite{Seiberg:1999xz,Maldacena:2000hw}, which describes fundamental strings propagating out to the \adsthree\ boundary in the background of electric and magnetic NS 3-form flux, is precisely such an instability.  This pathology can be avoided by turning on any of the four moduli (for instance the combination of $C_0$ and $C_4$ which preserves the fixed scalar condition) that take the theory away from the singular locus.  In the slice of the moduli space depicted in Figure~\ref{fig:cuspform}, there is a singular locus at $\Re(\tau)=0$ (the red dashed line) and at each of the images of this line under the maps that permute the weak-coupling cusps (in this example, the magenta dashed arc between $\tau=1/2$ and $\tau=1/3$).

The description of the CFT in terms of a sigma model on the moduli space of $n_1$ instantons in $U(n_5)$ gauge theory is an approximate weak-coupling description in a particular cusp, corresponding to a particular choice of factorization.  In the cusp where $n_5=1$ and $n_1=N$, there is a (codimension four) weak-coupling locus where the sigma model target space $\cX$ is the symmetric product orbifold%
~\cite{Vafa:1995zh,Bershadsky:1995qy} (see also the review~\cite{David:2002wn})
\be
\cX_0 = \bigl(\cM^N\bigr)/S_N
\ee
which is a solvable conformal field theory.
Note that the map~\eqref{dualityrotation} does \emph{not} imply that there is a symmetric orbifold description for every cusp; in fact it is rather unlikely that there is one.  The analysis in~\cite{Larsen:1999uk} of the masses of states carrying conserved charges in the ${\underline{\bf 16}}$ of branes and momenta on the $\IT^4$ showed that the energetics was consistent with the corresponding charges in the symmetric orbifold only if the latter was a weak-coupling limit in the cusp where $n_1=N$ and $n_5=1$.  The sigma model on the moduli space of instantons may be a weak-coupling description of other cusps, but it does not reduce to the symmetric product orbifold at low energies.

The symmetric orbifold is a nonsingular, parity-invariant CFT\@.  In the cusp corresponding to $n_5=1$ the parity-invariant points are at $C_0=0$ and $C_0=1/2$.  The former is the singular locus, which leaves $C_0=1/2$ as the orbifold locus.  The $SL(2,\IZ)$ map~\eqref{dualityrotation} from the cusp at $\tau=i\infty$, corresponding to the symmetric orbifold, to a cusp at $\tau\sim \frac{a}{n_1}$, with macroscopic charges $(n_1,n_5)$ (where the supergravity description is valid) has $a n_5- bn_1=1$.  
The cusp is a macroscopic distance in the natural hyperbolic metric $\frac{|d\tau|^2}{(\Im\tau)^2}$ from any point along the orbifold locus.

The regions of the moduli space admitting a low-energy supergravity description are distant from the solvable locus $\cX_0$, and hence it is not possible in general to relate states in the solvable CFT with particular supergravity backgrounds.  
Nevertheless for BPS states one can compare quantities such as conformal dimensions and three-point correlators, which are protected by supersymmetry against renormalization as we move across the moduli space~\cite{Baggio:2012rr}. In this section we provide a dictionary between the asymptotically-AdS geometries of Section~\ref{Sect:AdS} and particular CFT states in the RR sector of the orbifold CFT\@. This dictionary should be interpreted in the following sense: The three-point correlators between these RR states and any chiral primary operator can be calculated either holographically using the supergravity solutions, or at the orbifold point using the free-field realization of the CFT, and the two results match. This point of view was introduced in~\cite{Skenderis:2006ah} in the sector of the RR ground states that are dual to two-charge geometries, and was extended in~\cite{Giusto:2015dfa} to the three-charge geometries of~\cite{Bena:2015bea}.  Of course for non-protected quantities, such as four-point functions, the effects of wavefunction renormalization generically become visible and the relation between the gravity solutions and the orbifold CFT states described here becomes less useful. With this understood, we now identify and discuss the holographic dictionary; our notation and conventions mostly follow~\cite{Giusto:2015dfa,Bena:2016agb}.

\subsection{Dual states}
\label{Sect:dual-CFT-states}

The twisted-sector ground states of the symmetric orbifold $(\cM)^N/S_N$
CFT in the RR sector are \nBPS{4}, and map to known supergravity
supertube geometries~\cite{Lunin:2001fv,Kanitscheider:2007wq}.  There is
an independent twisted sector for each conjugacy class in the symmetric
group.  Symmetric group elements consist of \emph{words} which are
products of (non-overlapping) cyclic permutations of the copies of
$\cM$.  The conjugacy class of a word is characterized by the number
$N_k$ of cycles of length $k$ in the word, with the total length
(including cycles of length one) being $\sum_k kN_k = N$.

When $k$ copies of the CFT on $\cM$ are sewn together by a cyclic permutation boundary condition, the result can be thought of as the CFT on $\cM$ on the $k$-fold covering of the coordinate cylinder on which the CFT lives.  The supersymmetric ground states of the $k$-cyclic twisted sector are thus the same as those of $\cM$.  For $\cM=\IT^4$, these ground states consist of ultrashort multiplets labelled by spin-1/2 doublets $\alpha$, $\alphabar$ under the $SU(2)\times SU(2)$ $\cR$-symmetry, and $A,B$ under an auxiliary $SU(2)_\cA$:
\be
\label{R gd states}
\ket{\alpha\alphabar}_k~,\quad
\ket{A B}_k~,\quad
\ket{\alpha B}_k~,\quad
\ket{A\alphabar}_k~;
\ee
The highest-weight states of the first two of these multiplets are bosonic, while in the last two they are fermionic.  We will focus on two ground states in particular -- the highest-weight state $\ket{++}_k$ of the $\cR$-symmetry bispinor multiplet (the first one in \eqref{R gd states}), and the singlet combination of the auxiliary $SU(2)_\cA$ bispinor (the second one in \eqref{R gd states}),
\be
\ket{00}_k \equiv \epsilon^{AB}\ket{AB}_k~.
\ee
The full ground state is then a tensor product of ground states for the cyclic twists in the symmetric group conjugacy class, having $N_k^{(s)}$ copies of $k$-cycle ground states~\eqref{R gd states} of the polarization state $s$.  We often refer to the cycles of the symmetric product as \emph{`strands'} of the dual CFT\@.  The class of states we are interested in thus takes the form
\be
\label{numberstates}
\psi_{\{N_k^{s}\}}\equiv\prod_{k,s} \bigl( \ket{s}_{k}^{\strut}\bigr)^{N_k^{s}}  ~.
\ee

The role of the various polarizations of cyclic twist is illustrated by the map between the \nBPS{4} states and their dual geometries~\cite{Skenderis:2006ah,Kanitscheider:2006zf,Giusto:2015dfa}:
\begin{subequations}\label{generaltwocharge}
\begin{align}
& Z_2 = 1 + \frac{Q_5}{L} \int_0^{L} \frac{1}{|x_i -g_i(v')|^2}\, dv'~,~~~
  Z_4 = - \frac{Q_5}{L} \int_0^{L} \frac{\dot{g}_5(v')}{|x_i -g_i(v')|^2} \, dv' \,,\\\label{Z1profile}
& Z_1 = 1 + \frac{Q_5}{L} \int_0^{L} \frac{|\dot{g}_i(v')|^2+|\dot{g}_5(v')|^2}{|x_i -g_i(v')|^2} \, dv' ~, ~~~ d\gamma_2 = *_4 d Z_2~,~~~d\delta_2 = *_4 d Z_4~,\\
& A = - \frac{Q_5}{L} \int_0^{L} \frac{\dot{g}_j(v')\,dx^j}{|x_i -g_i(v')|^2} \, dv' ~, ~~~ dB = - *_4 dA~,~~~ ds^2_4 = dx^i dx^i~, \\
& \beta = \frac{-A+B}{\sqrt{2}}~,~~~\omega = \frac{-A-B}{\sqrt{2}}~,~~~{\cal F}=0~,~~~a_1=a_4=x_3=0~,
\end{align}
\end{subequations}
where the dot on the profile functions indicates a derivative with respect to $v'$, $L\equiv 2\pi Q_5/R_y$, 
and $*_4$ is the Hodge dual with respect to the flat $\mathbb{R}^4$ metric $ds^2_4=dx^i dx^i$.

One can expand the two-charge profile functions in Fourier series
\bea
g_1+ i g_2 &=& \sum_{\ell>0}\Bigl( \frac{a_\ell^{ +\! +}}{\ell} e^{\frac{2\pi i \ell}{L} v'}
+  \frac{a_\ell^{ --}}{\ell} e^{-\frac{2\pi i \ell}{L} v'} \Bigr)~,
\nn\\
g_3+ i g_4 &=& \sum_{\ell>0}\Bigl( \frac{a_\ell^{+-}}{\ell} e^{\frac{2\pi i \ell}{L} v'}
+  \frac{a_\ell^{ -+}}{\ell} e^{-\frac{2\pi i \ell}{L} v'} \Bigr)~,
\nn\\
g_5 &=& -{\rm Im}\Bigl[ \sum_{\ell>0}\frac{a_\ell^{00}}{\ell} e^{\frac{2\pi i \ell}{L} v'} \Bigr]  ~.
\eea
subject to the constraint on the overall amplitude
\be
\sum_\ell \Bigl( | a_\ell^{+\!+} |^2 + | a_\ell^{--} |^2 + | a_\ell^{+-} |^2 + | a_\ell^{-+} |^2 + | a_\ell^{00} |^2 \Bigr) = \frac{Q_1Q_5}{R_y^2} ~.
\ee
The specific solutions of Section~\ref{Sect:AdS} are built starting from the ground states
\be
\label{gmodeamplitudes}
a_1^{+\!+} \equiv a
\quad,\qquad
a_k^{00} \equiv b_k = b_4^{k,0,0}
\ee
with all other coefficients equal to zero.

As we see from~\eqref{generaltwocharge}, the numbers $N_k^i$ of cycles
with polarization $\sigma^i_{\alpha\alphabar}\ket{\alpha\alphabar}_k$ in
the number eigenstates~\eqref{numberstates} determine the amplitudes of
the Fourier coefficients of the functions $g_i(v)$ and thus specify
gyrations of the brane bound state in the four dimensions transverse to
its worldvolume.  Having only $\ket{++}_1$ strands corresponds to a round
supertube rotating in the $x_1$-$x_2$ plane.  The $\ket{00}_k$ strands
carry no transverse angular momentum, and so do not affect the shape of
the supertube.  Their numbers $N_k^{00}$ do however determine the
amplitudes of the Fourier coefficients of the function $g_5$ which
specifies the harmonic function $Z_4$ and therefore affects the
antisymmetric tensor fields of the supergravity
background.  Because the fields of the supergravity solution have both a
well-defined amplitude \emph{and} phase, they are represented as
coherent states built from the number eigenstates $\psi_{\{N_k^{(s)}\}}$
(see for instance equations (3.6)--(3.12) of~\cite{Giusto:2015dfa}).

The three-charge states dual to the geometries of Sections
\ref{Sect:first-layer} and \ref{Sect:AdS} are built on these unexcited
($m=n=0$) round supertubes.  The momentum-generating excitation labelled
by $m$ in supergravity adds $J_L$ charge and $P$ charge in equal
proportion to the harmonic function $Z_4$; one can identify it as
corresponding to the action of $J_{-1}^+$ on the $\ket{00}$ strands of
the \nBPS{4} ground state~\cite{Bena:2015bea}.  

Under spectral flow to
the NS-NS sector, $\ket{00}_k$ is mapped into an anti-chiral
primary state $\ket{00}_k^{\rm NS}$ with $h=-j^3=k/2$, and $J_{-1}^+$ is mapped to
$J_0^+$.  Because $\ket{00}_k^{\rm NS}$ is the lowest-weight state of
$SU(2)_L$, it can be acted on by $J_0^+$ a maximum of $k$ times, which means
that $m\le k$.  Similarly, the generalization to $n>0$ involves
additional CFT excitations which carry $n$ units of momentum but no
angular momentum; it is natural to identify them with the mode operator
$(L_{-1}-J_{-1}^3)$, which commutes with $J_{-1}^+$.  This discussion is
completely in parallel to the one we gave on the gravity side in Section
\ref{Sect:first-layer}.
Thus we
are led to the set of states
\begin{equation}
 \label{eq:o3cstate}
\psi_{\{N_1,N_{k,m,n}\}} 
\equiv \bigl(\ket{++}^{\strut}_1\bigr)^{N_1 } 
\prod_{k,m,n} \biggl(\frac{(J^+_{-1})^{m}}{m!} \frac{(L_{-1}- J^3_{-1})^n}{n!} |00\rangle_k\biggr)^{N_{k,m,n} }.
\end{equation}
This is the more precise version of the ``intuitive'' formula that we
presented in \eqref{CFT_state_w_multiple_modes}. The numbers $\{N_1,
N_{k,m,n}\}$ specify the number of strands with particular quantum numbers
and
must satisfy\footnote{On the supergravity side, this
constraint can be understood as the level-matching constraint on the
worldsheet of the F1-P supertube which is in the same duality orbit as
the \nBPS{4} D1-D5 supertube ground state.}
\begin{equation}
  \label{totwind1}
 N_1+ \sum_{k,m,n} k N_{k,m,n} = N~. 
\end{equation}
In \eqref{eq:o3cstate}, we considered only the ground state $\ket{00}_k$
with excitations on it, but we can in principle include all the other ground
states in \eqref{R gd states}.\footnote{The generalization of
\eqref{eq:o3cstate} and \eqref{totwind1} to include all ground states is
\begin{equation}
\psi_{\{N_{k,m,n}^s\}} 
\equiv
\prod_{k,m,n,s} \biggl(\frac{(J^+_{-1})^{m}}{m!} \frac{(L_{-1}- J^3_{-1})^n}{n!} |s\rangle_k\biggr)^{N_{k,m,n}^s },\qquad
 \sum_{k,m,n,s} k N_{k,m,n}^s = N~. 
\end{equation}
From this perspective, the numbers $N_1$ and $N_{k,m,n}$ in \eqref{eq:o3cstate}
should more consistently be denoted by $N_{1,0,0}^{++}$ and
$N_{k,m,n}^{00}$, respectively.  The
supergravity solutions dual to the more general states will have base space data,
$(\mathcal{B},\beta)$, that is more complicated than the base space used in this paper.  In
\cite{Bena:2016agb}, another set of special states for which the data $(\mathcal{B},\beta)$ remain simple (called ``Style 1'' states) are discussed.}

The classical supergravity dual does not correspond to the state
\eqref{eq:o3cstate} with fixed numbers $\{N_1,N_{k,m,n}\}$ but rather to
its coherent superposition
\cite{Skenderis:2006ah,Kanitscheider:2006zf,Kanitscheider:2007wq}.  We
introduce a set of dimensionless parameters $\{A_1,B_{k,m,n}\}$, which are
closely related to the supergravity mode amplitudes $a$ and $b_4^{k,m,n}$
of~\eqref{Z4Th4_solngen}.  The state dual to the
coiffured supergravity solution can be written, generalizing the $n=0$
expression in \cite{Giusto:2015dfa}, as
\begin{equation}
\label{eq:gravityCFTmap3c}
\psi({\{A_1,B_{k,m,n}\}}) = \sideset{}{'}\sum_{\{N_1,N_{k,m,n}\}}
A_1^{N_1} 
\Bigl[\,\prod_{k,m,n}(B_{k,m,n})^{N_{k,m,n}}\Bigr]
~\psi_{\{N_1,N_{k,m,n}\}}~,
\end{equation}
where the sum is restricted to $\{N_1,N_{k,m,n}\}$ satisfying
\eqref{totwind1}.  In the large $N$ limit this sum is dominated by a
stationary point $\{\Nb_1,\Nb_{k,m,n}\}$ which can be found by
calculating the norm $|\psi({\{A,B_{k,m,n}\}})|^2$ and taking its
variation with respect to $\{N_1,N_{k,m,n}\}$. In order to do this,
we need to derive the effect of the momentum-carrying perturbations
$J^+_{-1}$ and $(L_{-1}- J^3_{-1})$ on the normalization of the
state~\eqref{eq:o3cstate}.  For  $n=0$ the result is given
in equation~(3.17) of~\cite{Giusto:2015dfa} and the generalization to $n\ne0$ is
given in Appendix~\ref{Sect:Norms}\@.  Using the result, the saddle-point values are found to be
\begin{equation}
  \label{eq:g321}
  \overline{N}_1 = |A|^2~,
  \qquad
  k \overline{N}_{k,m,n} = {{k}\choose{m}} {{n+k-1}\choose{n}} |B_{k,m,n}|^2\;.
\end{equation}

Thus far, we have been considering the general set of states that have
strands with different quantum numbers $(k,m,n)$; namely, $N_{k,m,n}\neq 0$
for multiple sets of values $(k,m,n)$. Now, let us focus on the special
states \eqref{eq:o3cstate} where $N_{k,m,n}$ is non-zero only for one particular set of values $(k,m,n)$, 
which can be written as
\begin{equation}
 \label{eq:o3cstate2}
\psi\strut_{N_1,N_{k,m,n}} 
\equiv \bigl(\ket{++}^{\strut}_1\bigr)^{N_1 } 
\biggl(\frac{(J^+_{-1})^{m}}{m!} \frac{(L_{-1}- J^3_{-1})^n}{n!} |00\rangle_k\biggr)^{N_{k,m,n} }.
\end{equation}
In this expression, $k,m,n$ are not summed over, but are fixed
numbers.  The corresponding coherent state~\eqref{eq:gravityCFTmap3c} can be written in terms of two quantities $A_1,B_{k,m,n}$ as
\begin{equation}
\label{eq:gravityCFTmap3c2}
\psi(A_1,B_{k,m,n}) = \sideset{}{'}\sum_{N_1,N_{k,m,n}}
A_1^{N_1} (B_{k,m,n})^{N_{k,m,n}}\,\psi\strut_{N_1,N_{k,m,n}}~,
\end{equation}
where the two numbers $N_1,N_{k,m,n}$ satisfy
\begin{align}
\label{totwind2}
 N_1+kN_{k,m,n}=N~.
\end{align}
We propose that the states \eqref{eq:gravityCFTmap3c2} are the holographic duals of the single-mode
supergravity superstrata that we constructed in Section \ref{Sect:AdS}.
The saddle point values for $A_1,B_{k,m,n}$ are determined by
\eqref{eq:g321}.  If we substitute $N_1,N_{k,m,n}$ in
\eqref{totwind2} with their saddle point values, we obtain
\begin{align}
\label{totwind3}
 |A_1|^2+{{k}\choose{m}} {{n+k-1}\choose{n}} |B_{k,m,n}|^2
 =N.
\end{align}
If we compare this with \eqref{eq:Q1Q5}, we find that the 
dimensionless coefficients $A_1$, $B_{k,m,n}$ of the CFT are related
to the corresponding Fourier coefficients $a$ and $b_4^{k,m,n}$ in
supergravity via
\be
\label{amplmap}
|A_1| = R\sqrt{\frac{N}{Q_1Q_5}}\, a
\quad,\qquad
|B_{k,m,n}| = R\sqrt{\frac{N}{2Q_1Q_5}}\, {{k}\choose{m}}^{-1} {{n+k-1}\choose{n}}^{-1}  b_4^{k,m,n} ~.
\ee

The explicit proposal for the CFT states dual to the microstate geometries we constructed allows one to perform quantitative AdS/CFT studies that generalize those of \cite{Giusto:2015dfa}. We leave such an interesting investigation for future work.

\subsection{Comparison of conserved charges}
\label{Sect:comparison}

We can now compare the CFT parameters to those of the supergravity solutions.
From the expression for $Z_1$ in \eqref{generaltwocharge} we see that the D1 charge of the \nBPS{4} ground states is given by
\begin{equation}
 Q_1={Q_5\over L}\int_0^L \bigl(|\dot{g}_i(v')|^2+|\dot{g}_5(v')|^2\bigr)dv'.
\end{equation}
The supergravity charges $Q_1$, $Q_5$ are related to the quantized D1 and D5
numbers, $n_1$ and $n_5$, by
\begin{equation}
Q_1 = \frac{(2\pi)^4\,n_1\,g_s\,\alpha'^3}{V_4}
~~,~~~~ 
Q_5 = n_5\,g_s\,\alpha'\,,
\label{Q1Q5_n1n5}
\end{equation}
where $V_4$ is the coordinate volume of $T^4$.  
The relation between $Q_p$ and the quantized momentum number $n_p$ is
\be
Q_p = \frac{(2\pi)^4\,n_p\,g_s^2\,\alpha'^4}{V_4 R_y^2} = \frac{Q_1 Q_5}{R_y^2 N} n_p~.
\label{Qp_np}
\ee
The dimensionful angular momenta $J$, $\tilde{J}$ defined in
\eqref{beta_omega_and_J} are related to the quantized ones $j$, $\tilde
j$ by
\begin{equation}
J= \frac{(2\pi)^4 g_s^2 \alpha'^4}{V_4\,R_y}\, j = \frac{Q_1 Q_5}{R_y N}\, j \,,\quad \tilde J= \frac{(2\pi)^4 g_s^2 \alpha'^4}{V_4\,R_y}\, \tilde j = \frac{Q_1 Q_5}{R_y N}\, \tilde j\,.
\end{equation}

By using the dictionary between bulk and CFT quantities introduced in
the previous section it is possible to match the supergravity and CFT
calculations of the conserved charges and of the three-point functions
of chiral primary
operators~\cite{Lunin:2001fv,Kanitscheider:2007wq,Giusto:2015dfa}.
Here we focus on the conserved charges; these can be derived
by using the average number of each type of strands derived in
\eqref{eq:g321}. For instance, in the class of states we considered,
each strand of the type $\ket{00}_k$ carries $(m+n)$ units of momentum, thus the total momentum is equal to $(m+n)$ times the average number,
$\overline{N}_{k,m,n}$
\begin{equation}
  \label{eq:np}
n_p = (m+n) \overline{N}_{k,m,n} =  \frac{R_y^2 N}{Q_1 Q_5} \left[\frac{m+n}{2k}  {{k}\choose{m}}^{-1} {{n+k-1}\choose{n}}^{-1} (b_4^{k,m,n})^2 \right] .
\end{equation}
In the last step we used \eqref{eq:g321} and \eqref{amplmap} in order to show that the result matches perfectly~\eqref{eq:Qpsugra}. Similarly for the angular momenta, we find
\begin{equation}
  \label{eq:jljr}
\begin{aligned}
  j & =  \frac{1}{2} \overline{N}_{1} + m \overline{N}_{k,m,n} = 
 \frac{R_y^2 N}{2 Q_1 Q_5} \left[{a^2} + \frac{m}{k}  {{k}\choose{m}}^{-1} {{n+k-1}\choose{n}}^{-1} (b_4^{k,m,n})^2\right]\;,\\
  \tilde{j} & = \frac{1}{2} \overline{N}_{1}  =\frac{R_y^2 N}{2 Q_1 Q_5} a^2~,
\end{aligned}
\end{equation}
which exactly match the supergravity results in~\eqref{eq:Jsugra}.

\section*{Acknowledgments}

\vspace{-2mm}

We thank Samir Mathur, Harvey Reall and Jorge Santos for discussions.
The work of IB and DT was supported in part by John Templeton Foundation Grant
48222 and by the ANR grant Black-dS-String.
The work of SG was supported in part by the Padua University Project CPDA144437.
The work of EJM was supported in part by DOE grant
DE-SC0009924, and a FACCTS collaboration grant. 
The work of RR and MS was supported in part by the Science and Technology
Facilities Council (STFC) Consolidated Grants ST/P000754/1 and ST/L000415/1.
This work of MS was supported in part by JSPS KAKENHI Grant Numbers
16H03979, and MEXT KAKENHI Grant Numbers 17H06357 and 17H06359.
The work of DT was supported by a CEA Enhanced
Eurotalents Fellowship and a Royal Society Tata University Research Fellowship.
The work of NPW was supported by DOE grant DE-SC0011687.
For hospitality during the course of this work, EJM, MS and NPW are
grateful to the IPhT, CEA-Saclay; and EJM, DT and IB are grateful to the
Centro de Ciencias de Benasque.
SG, EJM, RR, MS, DT, and NPW thank the Yukawa Institute of Theoretical
Physics, Kyoto University for hospitality during the workshop ``Recent
Developments in Microstructures of Black Holes'' (YITP-T-17-05) where
this paper was completed.


\appendix

\section{Derivation of the explicit form of the function $F_{k,m,n}^{(p,q,s)}$}
\label{app:F_kmn}

In constructing the solution to the Layer 2 equations, 
one encounters the problem of finding the function
$F^{(p,q,s)}_{k,m,n}(r,\theta)$ satisfying
\begin{align}
 \widehat{\cL}^{(p,q,s)}F^{(p,q,s)}_{k,m,n}={\Delta_{k,m,n}\over (r^2+a^2)\cos^2\theta\,\, \Sigma} \,,
\label{LhatonFpqkmn}
\end{align}
where $\Delta_{k,m,n}$ and the scalar Laplacian with wave numbers $(p,q,s)$,
$\widehat{\cL}^{(p,q,s)}$, are defined in \eqref{Delta_v_kmn_def}
and \eqref{L0tand4defns} respectively.  In this appendix we derive the
explicit form of the solution $F^{(p,q,s)}_{k,m,n}(r,\theta)$.  In Section \ref{RMSsols}, we gave the explicit expression for $F_{2k,2m,2n}\equiv
F^{(0,0,0)}_{2k,2m,2n}$.  The derivation below is a
straightforward generalization of the derivation of $F_{k,m}^{(p,q)}$
done in Ref.~\cite{Bena:2015bea}.  For some intermediate steps that are not spelled out in the
derivation below, see Appendix B there.

Let us  first define
\begin{align}
 G_{k,m,n}={\Delta_{k,m,n}\over r^2+a^2}\,,\qquad
 S_{k,m,n}={\Delta_{k,m,n}\over (r^2+a^2)\cos^2\theta\, \Sigma}\,.
\end{align}
It is straightforward to  check that these functions satisfy the following recursion relation:
\begin{multline}
 \widehat\cL^{(p,q,s)}G_{k,m,n}
 =(n^2-s^2) S_{k+2,m+2,n-2}+((p+s)^2-(k+n+2)^2)S_{k+2,m+2,n}\\
 \qquad
 +((k-m)^2-(p-q)^2)S_{k,m+2,n}+(m^2-q^2)S_{k,m,n}\,.
\label{recursion_Gk,m,n}
\end{multline}
Introducing the generating functions
\begin{align}
\begin{split}
 \cF(\kappa,\mu,\nu)&\equiv \sum_{k,m,n}F^{(p,q,s)}_{k,m,n} \, e^{k\kappa+m\mu+n\nu},\\
 \cG(\kappa,\mu,\nu)&\equiv \sum_{k,m,n}G_{k,m,n}           \, e^{k\kappa+m\mu+n\nu},\\
 \cS(\kappa,\mu,\nu)&\equiv \sum_{k,m,n}S_{k,m,n}           \, e^{k\kappa+m\mu+n\nu},
\end{split}
\end{align}
we  can rewrite the equation we want to solve, \eqref{LhatonFpqkmn}, as
\begin{align}
 \widehat\cL^{(p,q,s)}\cF(\kappa,\mu,\nu)=\cS(\kappa,\mu,\nu)\,,
\end{align}
and the recursion relation
\eqref{recursion_Gk,m,n} as
\begin{multline}
 \widehat\cL^{(p,q,s)}\cG(\kappa,\mu,\nu)=
 \Bigl[e^{-2\kappa-2\mu+2\nu}((\partial_\nu+2)^2-s^2)+e^{-2\kappa-2\mu}((p+s)^2-(\partial_\kappa+\partial_\nu)^2)
 \\
 +e^{-2\mu}((\partial_\kappa -\partial_\mu +2)^2-(p-q)^2)+(\partial_\mu^2 -q^2)
 \Bigr] \cS(\kappa,\mu,\nu)\,.
\end{multline}
Since $\widehat\cL^{(p,q,s)}$ commutes with $\p_\kappa,\p_\mu,\p_\nu$,
the above equation means that
\begin{align}
 \cF
 &=
 -\Bigl[
 e^{-2\kappa-2\mu}((\partial_\kappa+\partial_\nu)^2-(p+s)^2)
 -e^{-2\mu}((\partial_\kappa -\partial_\mu +2)^2-(p-q)^2)
 \notag\\
 &
 \hspace{30ex}
 -(\partial_\mu^2 -q^2)
 -e^{-2\kappa-2\mu+2\nu}((\partial_\nu+2)^2-s^2)
 \Bigr]^{-1} \cG
 \notag\\[2ex]
 &=
 -\sum_{i=0}^\infty \Biggl[
 e^{2\kappa}{(\partial_\kappa -\partial_\mu +2)^2-(p-q)^2\over (\partial_\kappa +\partial_\nu +2)^2-(p+s)^2}
 +e^{2\kappa+2\mu}{\partial_\mu^2-q^2\over (\partial_\kappa +\partial_\nu +2)^2-(p+s)^2}
\notag\\
 &
 \hspace{10ex}
 +e^{2\nu}{((\partial_\nu+2)^2-s^2)\over (\partial_\kappa +\partial_\nu +2)^2-(p+s)^2}
 \Biggr]^i\,
  e^{2\kappa+2\mu}\,{1\over (\partial_\kappa +\partial_\nu +2)^2-(p+s)^2}\,
 \cG \,.
\end{align}
Expanding in a multinomial expansion and examining the coefficient of $e^{k\kappa+m\mu+n\nu}$, one finds:
\begin{align}
 F^{(p,q,s)}_{k,m,n}
 &=-{1\over 4}\sum_{i=0}^\infty \sum_{j_1+j_2+j_3=i}^\infty {i\choose j_1,j_2,j_3}
 \notag\\
 &\qquad
 \times
 \overbrace{(k_+ - m_+)(k_+ - m_+ -1)\cdots\,}^{j_1}\,
 \overbrace{(k_- - m_-)(k_- - m_- -1)\cdots\,}^{j_1}
 \notag\\
 &\qquad
 \times
 \overbrace{(m_+ - 1)(m_+ - 2)\cdots\,}^{j_2}\,
 \overbrace{(m_- - 1)(m_- - 2)\cdots\,}^{j_2}\,\,\,
 \overbrace{n_+ (n_+ - 1)\cdots\,}^{j_3}\,
 \overbrace{n_- (n_- - 1)\cdots\,}^{j_3}\,
 \notag\\
 &\qquad
 \times{1\over 
 \underbrace{(k_+ + n_+)(k_+ + n_+ -1)\cdots\,}_i\,
 \underbrace{(k_- + n_-)(k_- + n_- -1)\cdots\,}_i
 }
 \notag\\
 &\qquad
 \times
 G_{k-2(j_1+j_2+1),m-2(j_2+1),n-2j_3}
 \notag\\[1ex]
 &=-{1\over 4}\sum_{i=0}^\infty \sum_{j_1+j_2+j_3=i}^\infty {i\choose j_1,j_2,j_3}
 {(k_+ - m_+)!\over (k_+ - m_+ - j_1)}\, {(k_- - m_-)!\over (k_- - m_- - j_1)}
 \notag\\
 &\qquad
 \times
 {(m_+ - 1)!\over (m_+ - j_2 - 1)!}\, {(m_- - 1)!\over (m_- - j_2 - 1)!}\,\,
 {n_+!\over (n_+ - j_3)!}\, {n_-!\over (n_- - j_3)!}
 \notag\\
 &\qquad
 \times {(k_+ + n_+ - j_1 - j_2 - j_3 - 1)!\over (k_+ + n_+)!}\,
 {(k_- + n_- - j_1 - j_2 - j_3 - 1)!\over (k_- + n_-)!}
 \notag\\
 &\qquad
 \times
  G_{k-2(j_1+j_2+1),m-2(j_2+1),n-2j_3}
\end{align}
where
\begin{align}
{i\choose j_1,j_2,\dots,j_n}\equiv {i!\over j_1! j_2!\cdots j_n!},\qquad
 i=j_1+\dots+j_n
\label{def_multinom}
\end{align}
is the multinomial coefficient, and where we defined
\begin{align}
 k_\pm \equiv {k\pm p\over 2}\,,\qquad
 m_\pm \equiv {m\pm q\over 2}\,,\qquad
 n_\pm \equiv {n\pm s\over 2}\,.
\end{align}

In fact, the sum can be simplified because $\sum_{i=0}^\infty
\sum_{j_1+j_2+j_3=i}^\infty =\sum_{j_1,j_2,j_3=0}^\infty$.  Using the
definition \eqref{def_multinom}, we find that the explicit expression
for $F^{(p,q,s)}_{k,m,n}(r,\theta)$ is
\begin{align}
F^{(p,q,s)}_{k,m,n}
 &=-{1\over 4(k_+ + n_+)(k_- + n_-)}
 \sum_{j_1,j_2,j_3=0 }
{j_1+j_2+j_3\choose j_1,~ j_2,~ j_3}\notag\\
&\times
{{k_+ + n_+ - j_1 - j_2 - j_3 - 1 \choose k_+ - m_+ - j_1, ~ m_+ - j_2 - 1, ~ n_+ - j_3}
{k_- + n_- - j_1 - j_2 - j_3 - 1 \choose k_- - m_- - j_1, ~ m_- - j_2 - 1, ~ n_- - j_3}
\over 
{k_+ + n_+ - 1 \choose k_+ - m_+ , ~ m_+ - 1 ,~ n_+}
{k_- + n_- - 1 \choose k_- - m_- , ~ m_- - 1 ,~ n_-}
}
G_{k-2(j_1+j_2+1),m-2(j_2+1),n-2j_3}
\label{hrul8Sep16}
\end{align}
where the sum is over 
\begin{align}
j_1,j_2,j_3\ge 0,\qquad  j_1+j_2+j_3\le \min(k_+ + n_+,k_- + n_-)-1.
\end{align}

In particular, when $p=q=s=0$,
\begin{multline}
F^{(0,0,0)}_{2k,2m,2n}
 =-{1\over 4(k + n)^2}
 \sum_{j_1,j_2,j_3=0 }^{j_1+j_2+j_3\le k+n-1}
{j_1+j_2+j_3\choose j_1,~ j_2,~ j_3}
{{k + n - j_1 - j_2 - j_3 - 1 \choose k - m - j_1, ~ m - j_2 - 1, ~ n - j_3}^2
\over 
{k + n - 1 \choose k - m , ~ m - 1 ,~ n}^2
}\\
\times
G_{2(k-j_1-j_2-1),2(m-j_2-1),2(n-j_3)} \,.
\end{multline}

\section{Normalization of CFT states}
\label{Sect:Norms}

In this appendix, we compute the normalization of the CFT
states~\eqref{eq:o3cstate}.  Because the states we consider are obtained by exciting
the 1/4-BPS states \eqref{numberstates}, it is useful to recall the
norm of \eqref{numberstates}:
\be
\label{STnorm}
\cN_{\sst  \textstyle\rm ST} \equiv
|\psi_{\{N_k^s\}}|^2=
\frac{N!}{\prod_{k,s} N_{k}^s! \, k^{N_{k}^s}}~.
\ee
This given by the number of ways one can partition $N$ to obtain the desired distribution of strands; for details, see Section 3 (in particular Eq.\;(3.4)) of~\cite{Giusto:2015dfa}.

The normalizations of the excited states obtained by the action of
$J_{-1}^+$ and $(L_{-1}\!-\!J_{-1}^3)$ are determined in terms of those
of the ground state, \eqref{STnorm},  through the commutation relations of the ${\cal
N}=4$ superconformal algebra,
\begin{align}
[L_m,L_n] &= (m-n)L_{m+n} + \frac k2 \, m(m^2-1)\delta_{m,-n} \,, \nn\\
[J_m^a,J_n^b] &= i\epsilon^{abc} J^c_{m+n} + \frac k2 \, m\delta_{m,-n}\delta^{ab} \,, \\
[L_m,J^a_n] &= -n J^a_{m+n} \ ,\nn
\end{align}
where $k$ is the level of the $SU(2)$ current algebra and $c=6k$ is the Virasoro central charge.  On a strand of length $k$, the positive integer $k$ is indeed the level of the diagonal sum of the $k$ copies of the $\cN=4$ algebra being wound together by the $\IZ_k$ cyclic twist.  Define $J^\pm = J^1\pm iJ^2$ and consider the following state:
\begin{align}
J^-_1 (J^+_{-1})^m \ket{00}\strut_k
 =  \bigl(-2J^3_0+k + J^+_{-1}J^-_1\bigr)\; (J^+_{-1})^{m-1} \ket{00}\strut_k \;;
\end{align}
the $J^3_0$ operator evaluates to $m-1$ acting on the right.  Proceeding iteratively one arrives at
\begin{align}
J^-_1 (J^+_{-1})^m \ket{00}\strut_k 
 &= \Bigl[ -2\sum_{\ell=0}^{m-1}\ell+mk\Bigr]\, (J^+_{-1})^{m-1} \ket{00}\strut_k \nn\\
 &= m\bigl( k-(m-1) \bigr)\, (J^+_{-1})^{m-1} \ket{00}\strut_k \;.
\end{align}
Iterating this again for $(J_1^-)^m$ acting from the left, one finds
\begin{align}
\label{Jminusone}
\strut_k\bra{00} (J_1^-)^m \, (J^+_{-1})^m \ket{00}\strut_k &=
	m!\;\bigl(k-(m-1)\bigr)\bigl(k-(m-2)\bigr)\cdots\bigl(k\bigr) \nn\\
	&= m!\, \frac{k!}{(k-m)!} \;.
\end{align}
One finds similarly
\begin{align}
&\hspace{-4ex}
 (L_1-J^3_1)\,(L_{-1}-J^3_{-1})^n \ket{00}\strut_k \notag\\
	&= \Bigl(2L_0-2J^3_0+\frac k2 +(L_{-1}-J^3_{-1})(L_1-J^3_1)\Bigr)\, (L_{-1}-J^3_{-1})^{n-1}\ket{00}\strut_k \nn\\
	&= \Bigr[\bigl(2(n-1)+k \bigr)+\bigl(2(n-2)+k \bigr)+\dots + \bigl(2(0)+k\bigr)\Bigr]\, (L_{-1}-J^3_{-1})^{n-1}  \ket{00}\strut_k \nn\\
	&= n\bigl(k+(n-1)\bigr)\, (L_{-1}-J^3_{-1})^{n-1}  \ket{00}\strut_k ~;
\end{align}
once again iterating for the $n^{\rm th}$ power of the lowering operator one finds
\begin{align}
\label{Lminusone}
\strut_k\bra{00}(L_1-J^3_1)^n\,(L_{-1}-J^3_{-1})^n \ket{00}\strut_k 
	&= n!\,\bigl(k+(n-1)\bigr)\bigl(k+(n-2)\bigr)\cdots\bigl(k+(0)\bigr) \nn\\
	&= n!\, \frac{(k+n-1)!}{(k-1)!} ~.
\end{align}
Combining the results~\eqref{STnorm}, \eqref{Jminusone}, \eqref{Lminusone}, one finds the norm of the state~\eqref{eq:o3cstate}:
\begin{equation}
  \label{calN}
  |\psi_{\{N_1,N_{k,m,n}\}}|^2 
  = \frac{N!}{N_1!} \prod_{k,m,n}{1\over  N_{k,m,n}!} 
  \left[{1\over k}{{k}\choose{m}} {{n+k-1}\choose{n}}\right]^{N_{k,m,n}} \,.
\end{equation}

The classical supergravity dual does not correspond to the
state~\eqref{eq:o3cstate} but rather to its coherent
superposition
$\psi({\{A_1,B_{k,m,n}\}})$
given in
\eqref{eq:gravityCFTmap3c}.
The norm of this state is, using the results above,
\begin{align}
|\psi({\{A_1,B_{k,m,n}\}})|^2
 =
\sideset{}{'}\sum_{\{N_1,N_{k,m,n}\}}
&|A_1|^{2N_1} 
\Bigl[\,\prod_{k,m,n}|B_{k,m,n}|^{2N_{k,m,n}}\Bigr]
 \notag\\
& \times  \frac{N!}{N_1!} \prod_{k,m,n}{1\over  N_{k,m,n}!} 
  \left[{1\over k}{{k}\choose{m}} {{n+k-1}\choose{n}}\right]^{N_{k,m,n}}~,
\end{align}
where the sum is over $\{N_1,N_{k,m,n}\}$ satisfying the
constraint~\eqref{totwind1}.  In the large $N$ limit, the sum is
dominated by a stationary point $\{\Nb_1,\Nb_{k,m,n}\}$, which can be
obtained by setting to zero the variation with respect to
$\{N_1,N_{k,m,n}\}$ of the summand and using the Stirling formula.
The result is
\begin{equation}
  \label{eq:g321app}
  \overline{N}_1 = |A|^2~,~~~ k \overline{N}_{k,m,n} = {{k}\choose{m}} {{n+k-1}\choose{n}}\, |B_{k,m,n}|^2\;,
\end{equation}
which is a generalization of equation~(3.21) of~\cite{Giusto:2015dfa}.
The strand multiplicities $\{N_1,N_{k,m,n}\}$ are not independent
variables but satisfy the the constraint~\eqref{totwind1}. However this constraint
applies to the average values $\{\Nb_1,\Nb_{k,m,n}\}$ and so we have
\begin{equation}
  |A|^2 +  \sum_{k,m,n}{{k}\choose{m}} {{n+k-1}\choose{n}}\, |B_{k,m,n}|^2 = N\,.
\end{equation}


\newpage

\begin{adjustwidth}{-1mm}{-1mm} 

\bibliographystyle{utphys}      

\bibliography{microstates}       

\end{adjustwidth}


\end{document}